\documentclass{article}

\usepackage[english]{babel}

\usepackage[letterpaper,top=2cm,bottom=2cm,left=3cm,right=3cm,marginparwidth=1.75cm]{geometry}

\usepackage[backend=biber, style=nature,]{biblatex}
\usepackage{amsmath}
\usepackage{graphicx}
\usepackage[colorlinks=true, allcolors=blue]{hyperref}
\usepackage{authblk}
\usepackage{upgreek} 
\usepackage{hyperref}
\hypersetup{
    colorlinks=true,
    linkcolor=blue,
    filecolor=magenta,      
    urlcolor=cyan,
    pdftitle={Overleaf Example},
    pdfpagemode=FullScreen,
    }

\title{Predicting discrete-time bifurcations with deep learning}

\author[1*]{Thomas M. Bury}
\author[2]{Daniel Dylewsky}
\author[2]{Chris T. Bauch}
\author[3]{Madhur Anand}
\author[1]{Leon Glass}
\author[1$\dagger$]{Alvin Shrier}
\author[1$\dagger$]{Gil Bub}

\affil[1]{Department of Physiology, McGill University, 3655 Promenade Sir William Osler, Montreal, Quebec H3G 1Y6, Canada}
\affil[2]{Department of Applied Mathematics, University of Waterloo, Waterloo, Canada}
\affil[3]{School of Environmental Sciences, University of Guelph, Guelph, Canada}
\affil[$\dagger$]{Joint last authors}
\affil[*]{Corresponding author}

\date{}
\begin{document}
\maketitle

\begin{abstract}
Many natural and man-made systems are prone to critical transitions---abrupt and potentially devastating changes in dynamics.
Deep learning classifiers can provide an early warning signal (EWS) for critical transitions by learning generic features of bifurcations (dynamical instabilities) from large simulated training data sets.
So far, classifiers have only been trained to predict continuous-time bifurcations, ignoring rich dynamics unique to discrete-time bifurcations.
Here, we train a deep learning classifier to provide an EWS for the five local discrete-time bifurcations of codimension-1.
We test the classifier on simulation data from discrete-time models used in physiology, economics and ecology, as well as experimental data of spontaneously beating chick-heart aggregates that undergo a period-doubling bifurcation. 
The classifier outperforms commonly used EWS under a wide range of noise intensities and rates of approach to the bifurcation.
It also predicts the correct bifurcation in most cases, with particularly high accuracy for the period-doubling, Neimark-Sacker and fold bifurcations. 
Deep learning as a tool for bifurcation prediction is still in its nascence and has the potential to transform the way we monitor systems for critical transitions.
\end{abstract}

\section*{Introduction}

Many systems in nature and society possess critical thresholds at which the system undergoes an abrupt and significant change in dynamics \cite{scheffer2020critical,levin1998ecosystems}. In physiology, the heart can spontaneously transition from a healthy to a dangerous rhythm \cite{glass2020clocks}; in economics, financial markets can form a `bubble' and crash into a recession \cite{sornette2017stock}; and in ecology, ecosystems can collapse as a result of their interplay with human behaviour \cite{barlow2014modelling, henderson2016alternative}. These events, characterised by a sudden switch to a different dynamical regime, are referred to as critical transitions.

Critical transitions can be better understood with bifurcation theory \cite{kuznetsov1998elements, strogatz2018nonlinear}, a branch of mathematics that studies how dynamical systems can undergo sudden qualitative changes as a parameter crosses a threshold (a bifurcation). Many bifurcations are accompanied by critical slowing down---a diminishing of the local stability of the system---which results in systematic changes to properties of a noisy time series, such as its variance, autocorrelation and power spectrum \cite{wissel1984universal,wiesenfeld1985noisy,scheffer2009early}. 
These properties can be approximated analytically in the presence of different bifurcations \cite{kuehn2013mathematical,o2018stochasticity,bury2020detecting,wiesenfeld1985noisy}, and a corresponding observation in data can be used as an early warning signal (EWS) for the bifurcation \cite{scheffer2009early}. Systematic changes in variance and lag-1 autocorrelation have been observed prior to transitions in climate \cite{dakos2008slowing, boers2018early, boers2021observation}, geological \cite{hennekam2020early}, ecological \cite{pace2017reversal, wang2012flickering} and cardiac \cite{quail2015predicting} systems, suggesting the presence of a bifurcation. However, these EWS have limited ability in predicting the type of bifurcation \cite{kefi2013early, bury2020detecting} and can fail in systems with nonsmooth potentials \cite{hastings2010regime} or noise-induced transitions \cite{ditlevsen2010tipping}.

More recently, deep learning techniques have been employed to provide EWS for bifurcations \cite{bury2021deep, deb2022machine, dylewsky2022universal}. This involves training a neural network to classify a time series based on the type of bifurcation it is approaching, as well as appropriate controls \cite{bury2021deep, deb2022machine, dylewsky2022universal}. Unlike many applications of deep learning, this approach does not require abundant data from the study system, which, in the context of critical transitions, is often unavailable. (Unfortunately we do not have data from thousands or more ecosystems or climate systems that went through bifurcation.)
Instead, the approach generates a massive library of simulation data from generic models that possess each type of bifurcation. 
The neural network then learns generic features associated with each type of bifurcation, that can be recognised in an unseen time series of the study system.
This is enabled by the existence of universal properties of bifurcations that are manifested in time series as a dynamical system gets close to a bifurcation \cite{kuznetsov1998elements, wissel1984universal}.
In our previous work, we trained a deep learning classifier to provide an EWS for continuous-time bifurcations, and found it was effective at predicting transitions for real thermoacoustic, climate and geological transitions \cite{bury2021deep}.

Bifurcations can be partitioned according to whether they occur in continuous or discrete-time dynamical systems \cite{strogatz2018nonlinear, kuznetsov1998elements}. This distinction is important, since discrete-time dynamical systems (difference equations) can display very different behaviour to their continuous-time counterparts (differential equations).
As an example, consider the logistic model for population growth. When set up in continuous time (appropriate for populations with overlapping generations e.g. humans), the population grows smoothly as the reproduction rate increases. Whereas, when set up in discrete time (appropriate for populations with non-overlapping generations, e.g. insects), the population undergoes a sequence of period-doubling bifurcations to chaos \cite{may1974biological}. 
This involves transitioning from equilibrium dynamics to an oscillation of period 2, to an oscillation of period 4, 8, 16 etc. until the dynamics become chaotic. It is therefore important to develop EWS suitable for both continuous and discrete-time bifurcations. 
While indicators like variance and lag-1 autocorrelation can provide EWS for discrete-time bifurcations, the ability of a deep learning classifier at this task has not been investigated.

As well as in ecology, discrete-time bifurcations arise naturally in physiology \cite{glass2020clocks}, epidemiology \cite{allen1994some}, and economics \cite{westerhoff2008consumer}, where events can take place on a discrete timeline. To illustrate our approach, we will use model simulations from ecology, physiology and economics, as well as experimental data from spontaneously beating chick heart aggregates \cite{kim2009stochastic, quail2015predicting}. Following administration of a drug, in some aggregates the time interval between two heart beats begins to alternate i.e. there is a period doubling bifurcation (Figure 1). 
Such transitions can also occur for the human heart in the form of T-wave alternans, which increases a patient's risk for sudden cardiac death \cite{verrier2011microvolt}. The period-doubling bifurcation is accompanied by critical slowing down, so systematic changes in variance and lag-1 autocorrelation are expected and have been shown to provide an EWS in this system \cite{quail2015predicting}. The chick heart aggregates serve as a good study system to test the performance of EWS since we have multiple recordings, not all of which underwent a transition, allowing us to test for false positives.

\begin{figure}
    \centering
    \includegraphics[width=0.5\textwidth]{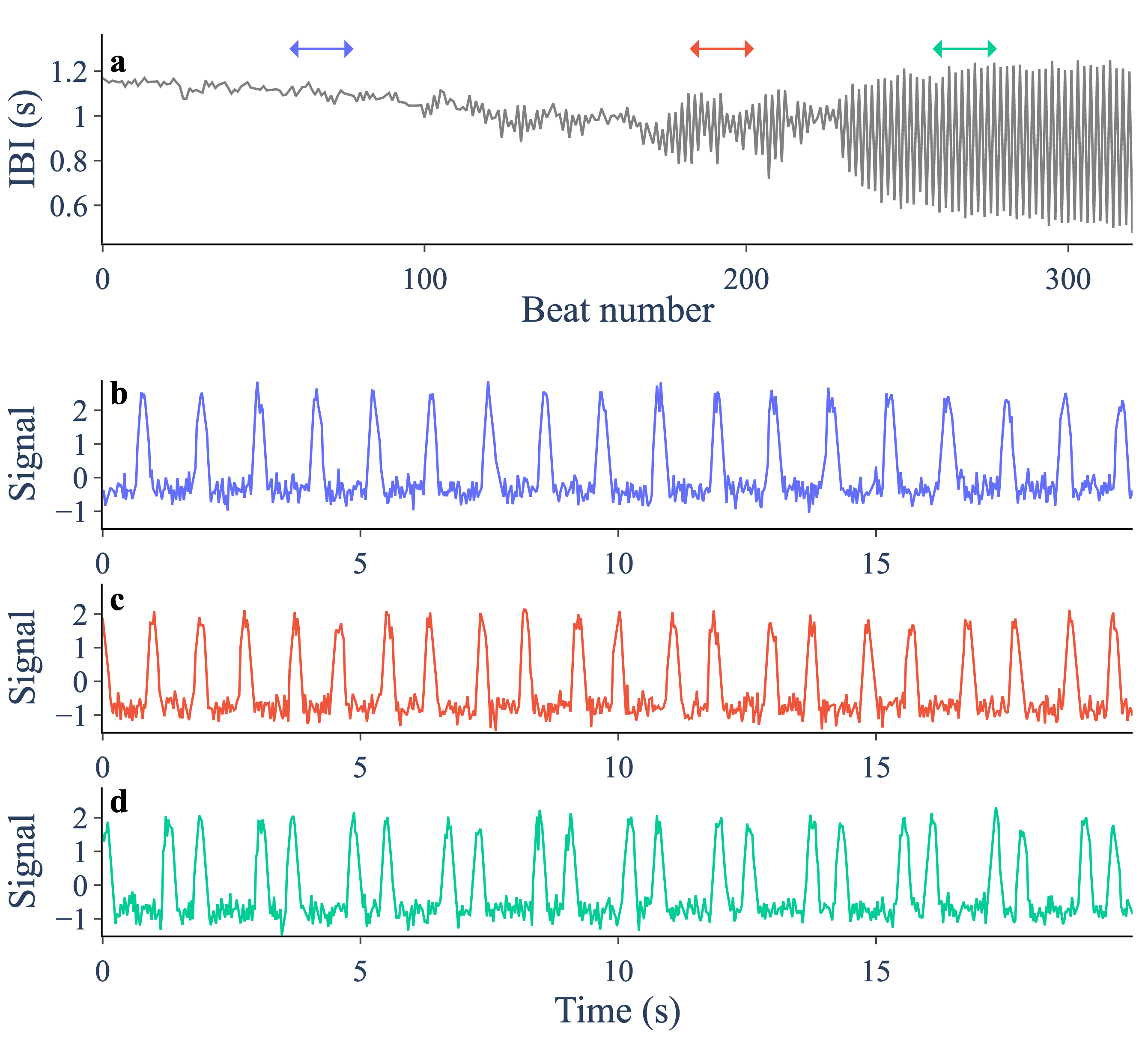}
    \caption{Period-doubling bifurcation in a spontaneously beating aggregate of embryonic chick heart cells following treatment with a potassium channel blocker (E-4031, 1.5$\upmu$mol). (a) Interbeat intervals (IBI) for consecutive beats. A period-doubling bifurcation occurs at approximately beat 230. Arrows correspond to traces plotted in lower panels. (b-d) Normalised signal from optical imaging of the aggregate's motion. Traces are from a section well before (b), just before (c), and after (d) the period-doubling bifurcation.}
    \label{fig:traj_select}
\end{figure}

Among discrete-time bifurcations, there are many types, each with an associated change in dynamics \cite{kuznetsov1998elements}. For this study, we focus on the five local bifurcations of codimension one (Box 1). In being `local', these bifurcations are accompanied by critical slowing down, so systematic changes and variance and autocorrelation are expected. However, not all of these bifurcations result in a critical transition \cite{kefi2013early}. They can instead involve a smooth transition to an intersecting steady state (transcritical) or to oscillations with gradually increasing amplitude (supercritical Neimark-Sacker). Predicting the type of bifurcation provides information on the nature of the dynamics following the bifurcation, something variance and autocorrelation alone do not provide.

Here, we train a deep learning classifier to provide a specific EWS for bifurcations of discrete-time dynamical systems. We train the classifier using simulation data of normal form equations appended with higher-order terms and noise. We then test the classifier on simulation runs of five discrete-time models used in cardiology, ecology and economics, and assess its performance relative to variance and lag-1 autocorrelation. We vary the noise amplitude and rate of forcing in model simulations to assess robustness of the EWS. Finally, we test the classifier on experimental data of spontaneously beating chick-heart aggregates that go through a period-doubling bifurcation. A reproducible run of all analyses may be performed on Code Ocean (\url{https://codeocean.com/capsule/2209652/tree/v1}) where the code is accompanied by the necessary software environment.

\begin{figure}[t!]
\fbox{
\begin{minipage}[t]{0.6\linewidth}
\raggedright
\setlength\parindent{16pt}
\noindent\textbf{Box 1. Local discrete-time bifurcations of codimension one.}\\
\noindent A discrete-time dynamical system has the form
\begin{equation}
    \vec{x}_{t+1} = f(\vec{x}_t)
\end{equation}
where $\vec{x}_t$ is a vector representing the state of the system and $f$ is a function that maps to the next state. The system is stable about a steady state $\vec{x}_*$ if all eigenvalues ($\lambda$) of the Jacobian matrix (gradient of $f$ at $\vec{x}_*$) have a magnitude less than one i.e. they lie within the unit circle on the complex plane (depicted in gray). A local bifurcation occurs when one or a pair of these eigenvalues (the dominant eigenvalue(s)) crosses the unit circle. The type of bifurcation depends on where the unit circle is crossed. A bifurcation is said to be of codimension one if it occurs upon changing only a single parameter of the system.

There are five types of local, discrete-time bifurcations of codimension one:\\
(a) period-doubling\\
(b) Neimark-Sacker\\
(c) fold\\
(d) transcritical\\
(e) pitchfork\\
\noindent Bifurcation diagrams are shown in the panels below \cite{kuznetsov1998elements}. The $x$-axis represents the bifurcation parameter and the $y$-axis a state variable. Solid lines show stable states or limit cycles, and dashed lines show unstable states. Arrows indicate direction of motion at different points in state space. The white points on the unit circle show the value of the dominant eigenvalue(s) at the bifurcation. The Neimark-Sacker bifurcation has a pair of complex conjugate eigenvalues, the angle ($\theta$) being the frequency of oscillations at the bifurcation.\\
\vspace{0.5cm}
\centering
\includegraphics[width=\linewidth]{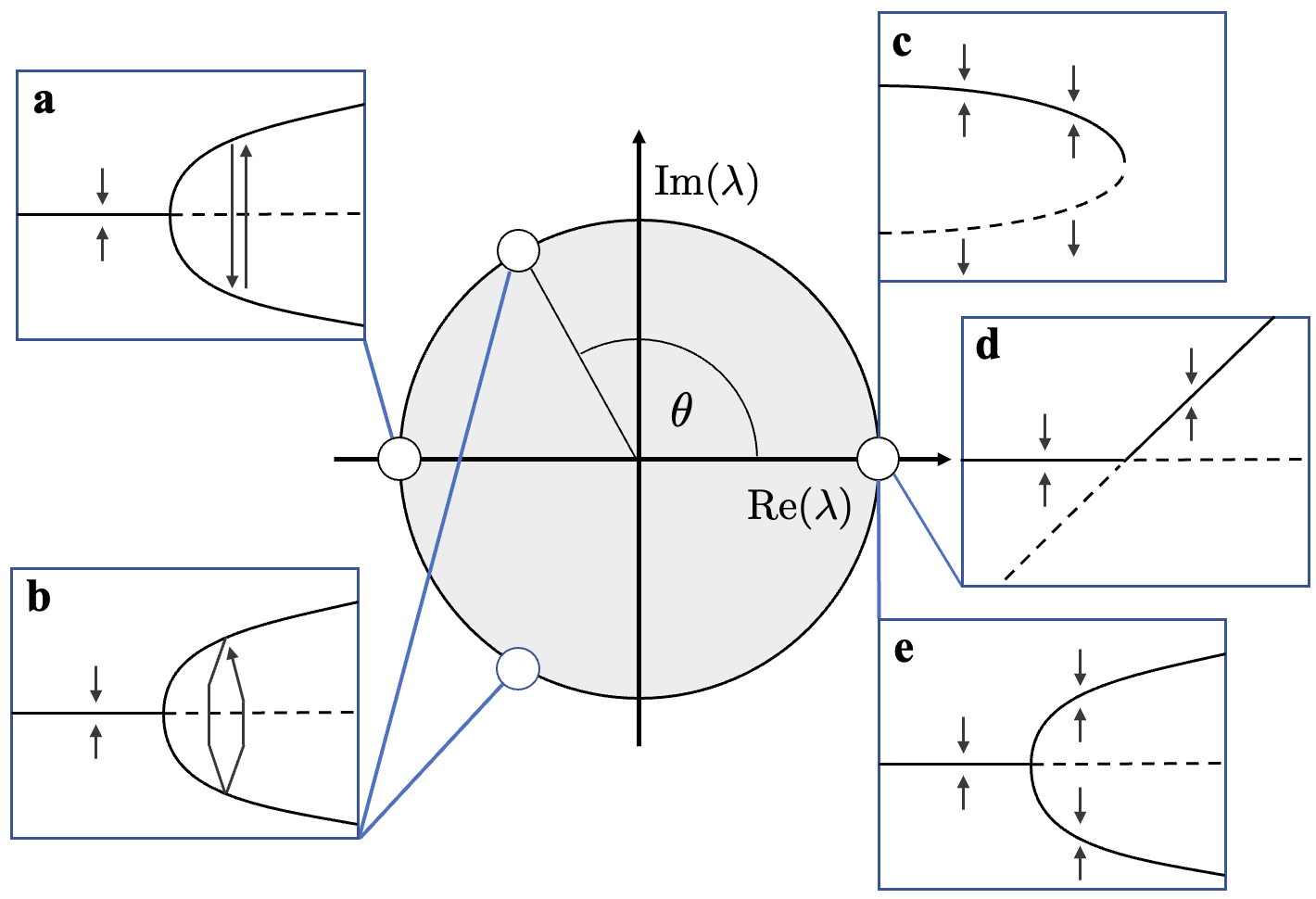}
\end{minipage}
}
\end{figure}

\section*{Results}

\textbf{Performance of classifiers on withheld test data}\\
As in previous work \cite{bury2021deep}, we train two different types of classifiers and use their ensemble average to make predictions. Classifier 1 is trained to recognise bifurcation trajectories based on middle portions of the time series, whereas Classifier 2 is trained on end portions (see Methods). In this way, Classifier 1 provides an earlier signal of a bifurcation and Classifier 2 provides a more specific signal, as more information is revealed closer to the bifurcation.
To quantify performance of the classifiers, we use the F1 score, which is a combined measure of sensitivity (how many of the true positives were predicted correctly) and specificity (how many of the positive predictions were actually true positives). On the withheld test data, Classifier 1 and 2 achieved an F1 score of $0.66$ and $0.85$, respectively. On the simpler, binary classification problem of predicting whether or not there will be any bifurcation, the classifiers achieved an F1 score of $0.79$ and $0.97$, respectively. Classifier 2 has a higher performance as it has the easier task of classifying data closer to the bifurcation where fluctuations are more pronounced. Performance on individual bifurcation classes is shown by confusion matrices (Sup. Fig.~1). The period-doubling, Neimark-Sacker and fold bifurcations are correctly classified with high sensitivity and specificity. On the other hand, the transcritical and pitchfork bifurcations are often mistaken for one another, likely due to having very similar normal forms (identical linear terms). Despite this, Classifier 2 can distinguish them at better than random, suggesting it is capable of recognising the different higher order terms in the data. From here onward, we report results using the ensemble prediction of the two classifiers, referred to collectively as the deep learning classifier.\\

\begin{figure}[t!]
    \centering
    \includegraphics[width=\textwidth]{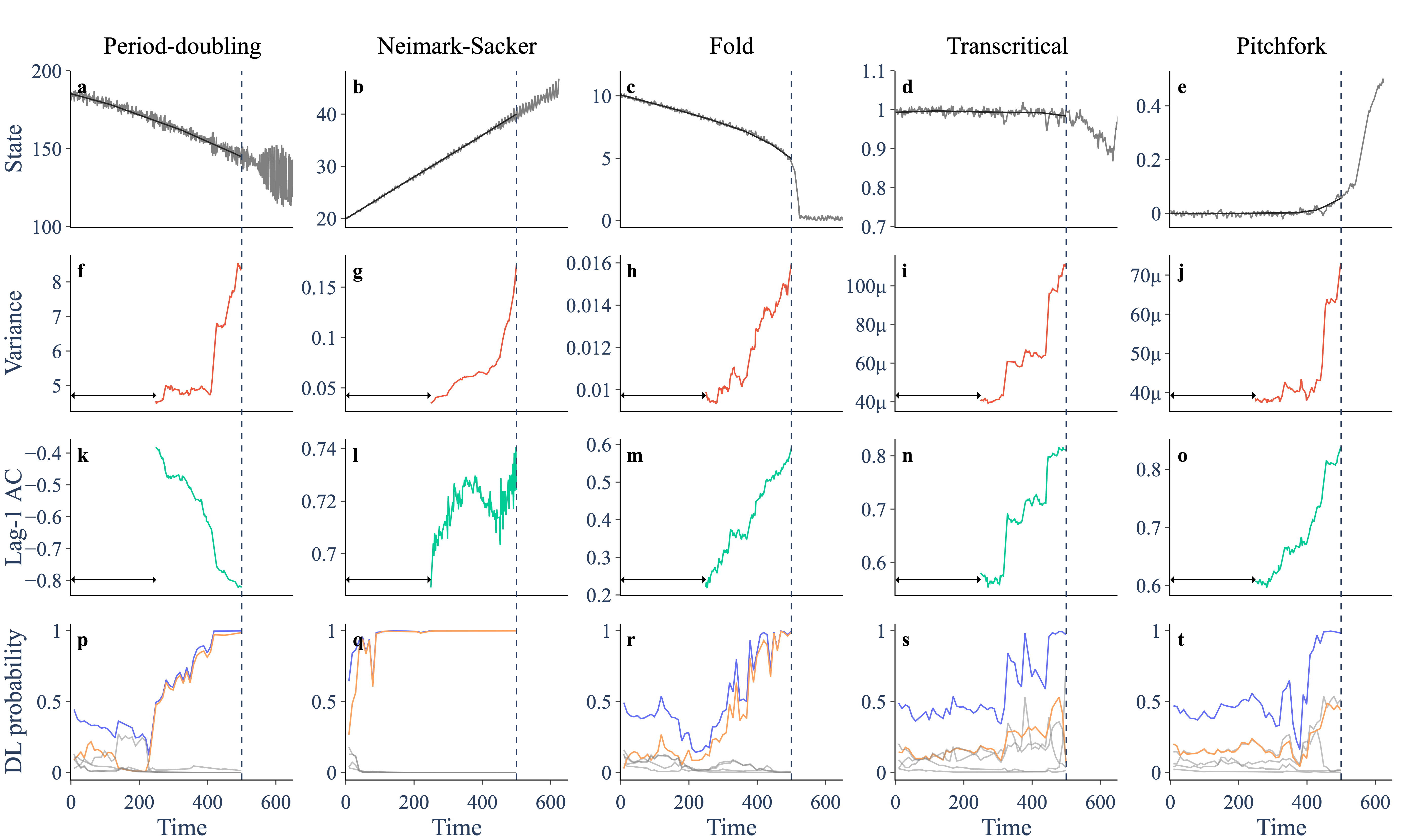}
    \caption{Trends in indicators prior to five different bifurcations in the theoretical models. (a-e) Trajectory (gray) and smoothing (black) of a simulation going through a period-doubling, Neimark-Sacker, fold, transcritical and pitchfork bifurcation, respectively. (f-j) Variance of residual dynamics after smoothing, computed over a rolling window (arrow) of size 0.5 times the length of the pre-transition data. (k-o) Lag-1 autocorrelation. (p-t) Probabilities assigned by the deep learning (DL) classifier when given all preceding data. Orange line shows probability assigned to the true bifurcation. Gray lines show probabilities assigned to the other bifurcations. Blue line shows the sum of the five bifurcation probabilities.}
    \label{fig:model_ews}
\end{figure}

\noindent\textbf{Performance of EWS on theoretical models}\\
We monitor variance, lag-1 autocorrelation and the deep learning classifier as progressively more of the time series is revealed. Variance and lag-1 autocorrelation are considered to provide an EWS if they display a strong trend, which we quantify using the Kendall tau statistic. For each of the five theoretical models (Fig.~\ref{fig:model_ews}(a-e)), we observe an increasing trend in variance (Fig.~\ref{fig:model_ews}f-j), and an increasing or decreasing trend in lag-1 autocorrelation (Fig.~\ref{fig:model_ews}k-o). The direction of the trend in lag-1 autocorrelation prior to a bifurcation depends on the frequency of oscillations ($\theta$) at the bifurcation---equivalently the angle of the dominant eigenvalue from the positive x-axis (Box 1). For $\theta\in[0,\pi/2)$ lag-1 autocorrelation increases, whereas for $\theta\in(\pi/2, \pi]$ it decreases---insights that can be obtained from analytical expressions of the autocorrelation function \cite{bury2020detecting,wiesenfeld1985noisy}. The period-doubling bifurcation is characterised by $\theta=\pi$, and the Neimark-Sacker bifurcation shown here has $\theta\approx\pi/4$. The trends in variance and lag-1 autocorrelation therefore behave as expected and can be used as an EWS.

The deep learning classifier assigns a probability to each of the six possible outcomes (null, period-doubling, fold, Neimark-Sacker, transcritical and pitchfork). It is considered to provide an EWS when there is a heightening in the sum of the bifurcation probabilities (blue line, Fig.~\ref{fig:model_ews}p-t). The type of bifurcation predicted is then taken as the highest individual bifurcation probability. 
For each simulation, the classifier is becomes more confident of an approaching bifurcation as time goes on, and its assigned bifurcation probability for the true bifurcation increases. The period-doubling, Neimark-Sacker and fold bifurcations are identified with high confidence well before the transition. The transcritical and pitchfork bifurcations are assigned roughly equal probability on their respective time series, suggesting they are difficult to tell apart---an observation consistent with the classifier's performance on its within-sample test data.

To obtain a measure of performance for the EWS, we need to test their predictions on both `forced' time series (where a bifurcation is approached) and `null' time series (where no bifurcation is approached). For the theoretical models, we generate null time series by keeping the bifurcation parameter fixed. We also test the robustness of the EWS to the rate of forcing and the noise amplitude of the simulations---two factors that have been shown to influence the performance of variance and lag-1 autocorrelation as an EWS \cite{clements2016rate, pavithran2021effect}. To this end, we simulate 100 forced and null time series at five different values of noise intensity and five different values of rate of forcing, resulting in a total of 5,000 time series for each theoretical model. Sample trajectories illustrating the different noise amplitudes and rates of forcing are shown in Sup. Fig.~3. We compute the probabilities assigned by the classifier and the Kendall tau values for variance and lag-1 autocorrelation at $80\%$ of the way through the pretransition time series, and use these values as discrimination thresholds to construct ROC curves (Fig.~\ref{fig:roc_combined}a-e). Using the AUC score (area under the ROC curve) as a measure of performance, we find that the classifier outperforms variance and lag-1 autocorrelation for each theoretical model. When evaluated for each combination of noise amplitude and rate of forcing separately, the classifier has the highest AUC score in $100\%$ of cases for the Neimark-Sacker, fold, and pitchfork models, $84\%$ of cases for the period-doubling model, and $80\%$ of cases for the transcritical model (Sup. Fig.~4). Similar to variance and lag-1 autocorrelation, the performance of the classifier is lower at higher rates of forcing. Noise amplitude affects performance differently depending on the model. In terms of predicting the correct bifurcation, the classifier typically performs better at slower rates of forcing (Sup. Fig.~5) and was able to classify the period-doubling and Neimark-Sacker bifurcations to high accuracy at all noise amplitudes and rates of forcing considered.\\

\begin{figure}[t!]
    \centering
    \includegraphics[width=\textwidth]{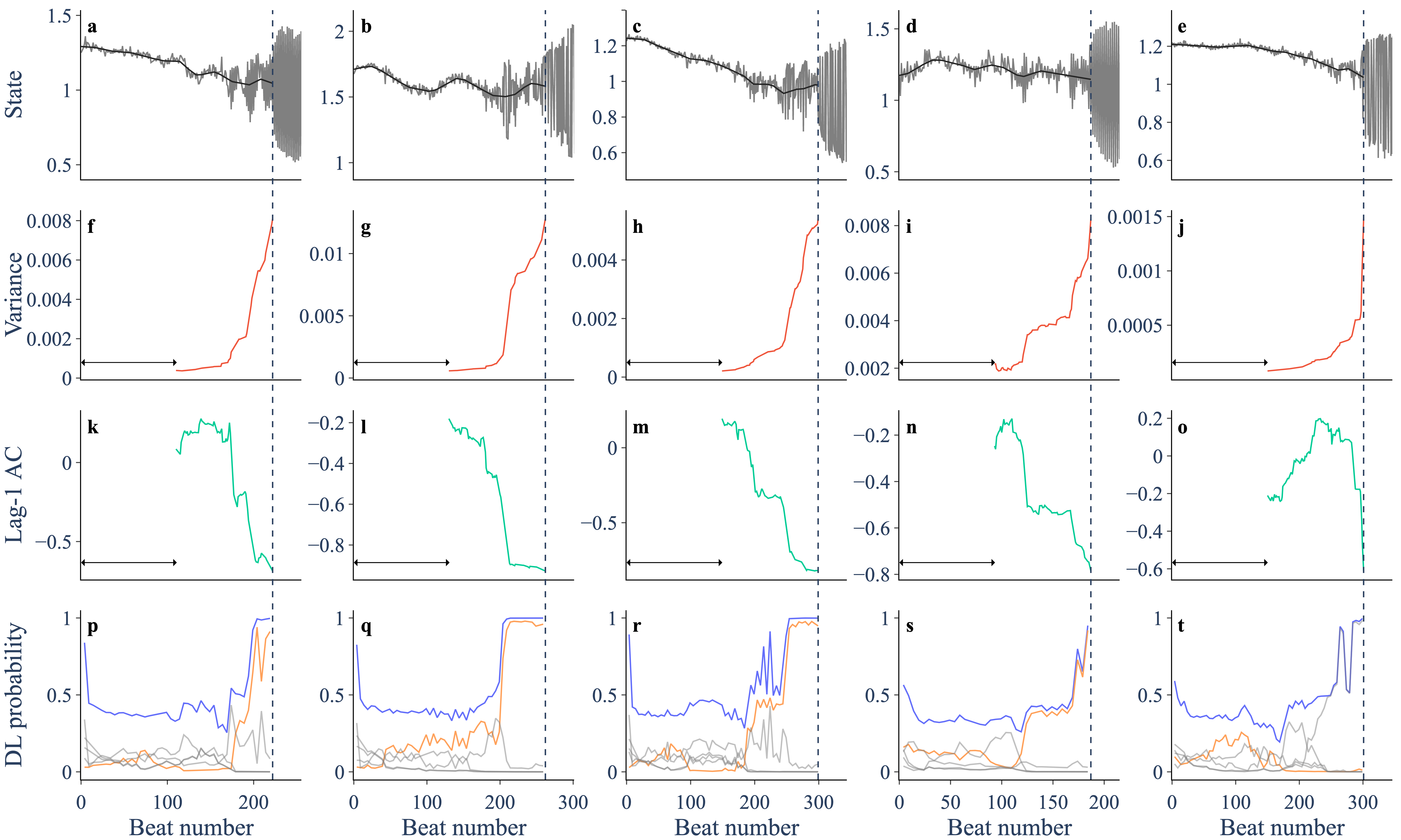}
    \caption{Trends in indicators prior to a period-doubling bifurcation in five chick heart aggregates treated with a potassium channel blocker. (a-e) Inter-beat interval (gray) and smoothing. (f-j) Variance of residual dynamics after smoothing, computed over a rolling window (arrow) of size 0.5 times the length of the pre-transition data. (k-o) Lag-1 autocorrelation. (p-t) Probabilities assigned by the deep learning (DL) classifier to the period-doubling bifurcation (orange) and the other bifurcations (gray). Blue line shows the sum of the five bifurcation probabilities.}
    \label{fig:heart_ews}
\end{figure}

\noindent\textbf{Performance of EWS on chick heart data}\\
In the chick heart data, we mostly observe an increasing trend in variance and a decreasing trend in lag-1 autocorrelation prior to the period-doubling bifurcation, as previously reported \cite{quail2015predicting}. This is consistent with analytical approximations for variance and lag-1 autocorrelation prior to a period-doubling bifurcation in a noisy dynamical system \cite{bury2020detecting, wiesenfeld1985noisy}. The 46 records and their EWS are shown in Sup. Fig.~6-9, and a sample of five period-doubling records are shown in Fig.~\ref{fig:heart_ews}. The classifier correctly predicts a period-doubling bifurcation in 16 of the 23 period-doubling records. In other cases, it incorrectly predicts a Neimark-Sacker bifurcation (e.g. Fig.~\ref{fig:heart_ews}e). This seems to be linked to an early increase in lag-1 autocorrelation, perhaps caused by a non-monotonic approach to the period-doubling bifurcation.
For predictions made at 60-100\% of the way through the chick heart data, the classifier obtains the highest AUC score (Fig.~\ref{fig:roc_combined}f), a slight improvement on variance, with the advantage of also providing the bifurcation type.

\begin{figure}[t!]
    \centering
    \includegraphics[width=\textwidth]{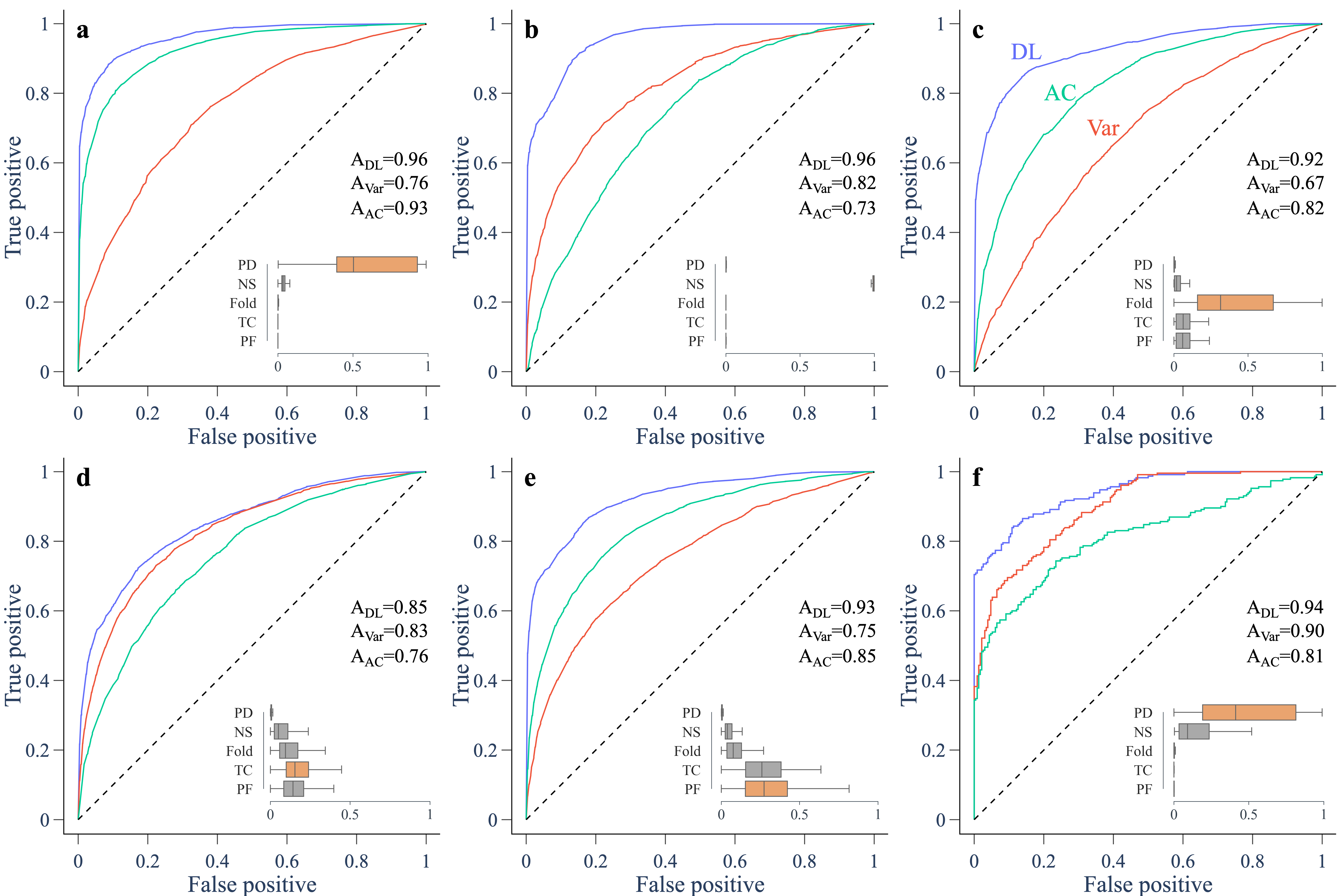}
    \caption{ROC curves for predictions of an upcoming transition in model and experimental data. ROC curves compare the performance of the deep learning classifier (DL, blue), variance (Var, red) and lag-1 autocorrelation (AC, green). For models (a-e), performance is assessed on 5,000 simulations with different noise amplitudes and rates of forcing. For experimental data (f), performance is assessed on 46 experimental runs. The area under the curve (AUC), abbreviated to A, is a measure of performance. Insets show the probabilities assigned by the classifier to each type of bifurcation (orange being the true bifurcation) among the trajectories approaching a transition. (a) Fox model going through a period-doubling bifurcation \cite{fox2002period}. (b) Westerhoff model going through a Neimark-Sacker bifurcation \cite{westerhoff2008consumer}. (c) Ricker model going through a fold bifurcation \cite{ricker1954stock}. (d) Lotka-Volterra model going through a transcritical bifurcation \cite{smith1968mathematical}. (e) Lorenz model going through a pitchfork bifurcation \cite{lorenz1989computational}. (f) Chick heart aggregates going through a period-doubling bifurcation \cite{kim2009stochastic}. Predictions are made 80\% of the way through the pretransition data for the models, and 60-100\% of the way for the experimental data. PD: period-doubling. NS: Neimark-Sacker. TC: transcritical. PF: pitchfork.}
    \label{fig:roc_combined}
\end{figure}

\section*{Discussion}

Many systems that evolve on a discrete timeline can undergo a sudden change in dynamics via a discrete-time bifurcation. We have found that a deep learning classifier is an effective tool for predicting discrete-time bifurcations in systems with a range of noise levels and rates of approach to the bifurcation. The classifier provides better sensitivity and specificity than variance and lag-1 autocorrelation---two commonly used EWS for bifurcations. Moreover, the classifier provides early indication of the type of bifurcation---an important piece of information given the qualitatively different dynamics associated with each bifurcation. A reliable early warning signal that specifies the type of bifurcation will help us prevent harmful bifurcations (e.g. dangerous heart rhythms \cite{glass2020clocks}) and promote favourable transitions (e.g. ecosystem recovery \cite{clements2019early}).

It may be possible to design a deep learning classifier that achieves a higher performance on our test data. First, there are many neural network architectures that could be investigated. For example, transformers, which are the current state-of-the-art for language models like GPT \cite{vaswani2017attention}, may also be useful for time series classification \cite{wen2022transformers}. Second, the hyperparameters of the classifier could be systematically tuned to optimise performance. Third, there may be benefit to reframing bifurcation prediction as a hierarchical classification problem \cite{silla2011survey}. One classifier could address the binary problem of flagging an approaching bifurcation, and a second classifier could address the multi-class problem of classifying the type of bifurcation \emph{given} that a bifurcation is approaching. (The same way one might want to distinguish images of dogs and cats before attempting to classify dog breeds.)

For a classifier to be effective, it must be trained on sufficiently diverse training data. As such, the method by which training data is obtained needs careful consideration. Our previous work on continuous-time bifurcations obtained training data from randomly generated dynamical systems with polynomial terms \cite{bury2021deep} and labelled the data using the bifurcation continuation software AUTO. This approach is appealing as it imposes relatively few restrictions on the models that are generated, and may include features associated with higher-order terms. Here, we opted for a more restricted approach that uses normal form models to generate the training data. This method has the advantage of being faster computationally, since the bifurcation of the model is known a priori.
This study shows that even with this more restricted training data, a classifier can generalise to detecting bifurcations in more complex model and empirical systems.

 We trained a classifier to provide EWS for a subset of bifurcations, namely local, codimension-one discrete-time bifurcations. While these bifurcations are present in many systems of interest, the real world presents many other classes of bifurcation in both continuous and discrete-time, including global bifurcations (e.g. homoclinic and heteroclinic), codimension-two bifurcations (e.g. cusp and Bogdanov-Takens), and bifurcations of attractors. For systems on attractors that explore a large portion of their phase space, empirical dynamical modelling \cite{ye2015equation}, reservoir computing \cite{patel2023using, kong2021machine} and deep neural networks \cite{lapeyrolerie2021teaching} can be used to make forecasts that may help predict critical transitions. In cases where spatial information is available, concepts from statistical physics may be useful \cite{hagstrom2021phase}, particularly in combination with deep learning \cite{dylewsky2022universal}. 

One limitation of the present classifier is that it is only trained to predict discrete-time bifurcations.
Therefore, one needs to know ahead of time whether continuous or discrete time is a better description for the system. 
In the case of the chick heart cells, we had prior knowledge that they are well described by a discrete-time dynamical system \cite{quail2012chaotic}, and therefore appropriate for the classifier. An interesting avenue for future research is to build a classifier that works for both continuous and discrete-time bifurcations. This may be achieved by generating a training library from models with a range of discretised timesteps, from very large steps that generate discrete-time bifurcations, down to the limit of a discrete timestep of zero, where continuous-time bifurcations occur. With a large enough training set, one would not need to assume ahead of time whether continuous or discrete time is a better description for the system.

Our results demonstrate that combining dynamical system and deep learning methodologies can provide EWS for critical transitions that are both more reliable and more descriptive than non-hybrid approaches. This study has set a baseline for prediction performance across a variety of popular discrete-time models and an experimental data set. In providing a code capsule that reproduces this study, we hope to facilitate the development, testing and comparison of related methods. Building a universal predictor for critical transitions is not a job for a single research team \cite{lapeyrolerie2021teaching}, and will benefit from a variety of approaches and open source code. Depending on context, critical transitions can be devastating or highly desirable. Improved EWS would allow us to better prevent or promote such transitions.

\section*{Methods}

\textbf{Generation of training data for the deep learning classifier}\\
Training data consists of simulation data from a library of 50,000 models. The models are generated at random from five different model frameworks, each possessing one of the bifurcations studied (period-doubling, Neimark-Sacker, fold, transcritical, pitchfork). The models are composed of the normal form of the bifurcation \cite{kuznetsov1998elements}, higher-order polynomial terms up to degree 10 with coefficients drawn from a normal distribution, and additive Gaussian white noise ($\epsilon_t$) with amplitude ($\sigma$) drawn from a uniform distribution. In each case, the bifurcation occurs at $\mu=0$.

The model for the period-doubling bifurcation is
\begin{equation}
    x_{t+1} = -(1+\mu)x_t  \pm  x_t^3  + \sum_{i=4}^{10} \alpha_i x_t^i + \sigma \epsilon_t,
\end{equation}
where $\alpha_i\sim \mathcal{N}(0,1)$. The positive (negative) cubic term yields a supercritical (subcritical) bifurcation, and is chosen at random.
The model for the Neimark-Sacker bifurcation is
\begin{equation}
    \begin{pmatrix}x_{t+1}\\y_{t+1}\end{pmatrix}
    = (1+\mu) R(\theta)
    \begin{pmatrix}x_t\\y_t \end{pmatrix} \pm (x_t^2 + y_t^2) R(\theta)
    \begin{pmatrix} x_t\\y_t \end{pmatrix}
     + \sum_{i=4}^{10} \sum_{j=0}^i
    \begin{pmatrix}\alpha_{ij}\\ \beta_{ij} \end{pmatrix}
    x_t^{i-j}y_t^j + 
    \sigma \begin{pmatrix} \epsilon_{t}^{(1)} \\ \epsilon_{t}^{(2)}  \end{pmatrix}
    \label{eq:ns}
\end{equation}
where $\alpha_{ij}$, $\beta_{ij}\sim\mathcal{N}(0,1)$, $R(\theta)$ is the rotation matrix
\begin{equation}
    R(\theta) = \begin{pmatrix} \cos \theta & -\sin \theta \\ \sin \theta & \cos \theta \end{pmatrix},
\end{equation}
and $\theta\sim\mathcal{U}[0,\pi]$ is the angular frequency of oscillations at the bifurcation. The positive (negative) cubic term yields a subcritical (supercritical) bifurcation, and is chosen at random. The model for the fold bifurcation is
\begin{equation}
    x_{t+1} = -\mu + x_t - x_t^2 + \sum_{i=3}^{10}\alpha_i(x_t-\sqrt{-\mu})^i +  \sigma \epsilon_t,
\end{equation}
where $\alpha_i\sim \mathcal{N}(0,1)$. 
The model for the transcritical bifurcation is
\begin{equation}
    x_{t+1} = (1+\mu)x_t  - x_t^2 + \sum_{i=3}^{10} \alpha_i x_t^i+ \sigma \epsilon_t,
\end{equation}
 where $\alpha_i\sim \mathcal{N}(0,1)$.
Finally, the model for the pitchfork bifurcation is
\begin{equation}
    x_{t+1} = (1+\mu)x_t  \pm  x_t^3 + \sum_{i=4}^{10} \alpha_i x_t^i + \sigma \epsilon_t,
\end{equation}
where $\alpha_i\sim \mathcal{N}(0,1)$. The positive (negative) cubic term yields a subcritical (supercritical) bifurcation, and is chosen at random.

The library is composed of 10,000 models from each framework. For each model, we run a `forced' simulation where the bifurcation parameter $\mu$ is increased linearly across the interval $[\mu_0,0]$, and a `null' simulation where $\mu$ is fixed at $\mu_0$. The initial value for the bifurcation parameter $\mu_0$ is drawn from a uniform distribution across all values that correspond to $|\lambda|<0.8$, where $\lambda$ is the eigenvalue of the Jacobian matrix in the model. This ensures that the training data contains simulations that start close to and far from a bifurcation. For the period-doubling, Neimark-Sacker, transcritical and pitchfork models, this means drawing $\mu_0$ from $\mathcal{U}[-1.8,-0.2]$. For the fold bifurcation, this means drawing $\mu_0$ from $\mathcal{U}[-0.9,-0.1]$. After a burn-in period of 100 iterations, we simulate each model for 600 iterations and keep the last 500 data points, or the 500 data points immediately preceding a transition if one occurs. We define a transition as a time when the deviation from equilibrium exceeds ten times the noise amplitude $\sigma$. We simulate one forced and one null simulation from each model, resulting in $50,000$ forced and $50,000$ null trajectories. To balance the number of entries for each class, we take $10,000$ null simulations at random, resulting in a total of $60,000$ entries in the training data set. Example trajectories for each class are shown in Sup. Fig.~2.

\vspace{0.5cm}
\noindent\textbf{Architecture and training of the deep learning classifier}\\
We use a neural network with a CNN-LSTM architecture and hyperparameters as in \cite{bury2021deep}. This consists of a single convolutional layer with max pooling followed by two LSTM layers with dropout followed by a dense layer that maps to a vector of probabilities over the six possible classes. For training, we use Adam optimisation with a learning rate of $0.0005$, a batch size of $1024$, and sparse categorical cross entropy as the loss function. We use a training/validation/test split of $0.95/0.025/0.025$. We found 200 epochs was sufficient to obtain optimal accuracy on the validation set.

To expose the classifier to time series of different lengths, we censor each time series in the training data. We train two classifiers independently using different censoring techniques. Classifier 1 is trained on time series censored at the beginning and the end, forcing it to learn from data in the middle of the time series. Classifier 2 is trained on time series only censored at the beginning, allowing it to learn from data right up to the bifurcation. The length for each censored time series $L$ is drawn from $\mathcal{U}[50,500]$. Then, for Classifier 1, the start time of the censored time series $t_0\sim U[0,500-L]$ and for Classifier 2, $t_0=500-L$. The censored time series are then normalised by their mean absolute value and prepended with zeros to make them $500$ points in length. We report results using the average prediction of the two classifiers.

\vspace{0.5cm}

\noindent\textbf{Theoretical models}\\
To test the deep learning classifier on out-of-sample data, we simulate a variety of nonlinear, discrete-time models, each containing one of the studied bifurcations. To account for stochasticity, we include additive Gaussian white noise. We run forced simulations, where the bifurcation parameter is increased linearly up to the bifurcation point, and null simulations, where the bifurcation parameter remains fixed. To create a diverse set of test data, we vary the noise amplitude ($\sigma$) and rate of forcing (rate of change of the bifurcation parameter). We run 100 forced and null simulations at five different noise amplitudes and five different rates of forcing, resulting in 5000 simulations of each model. Values for the noise amplitude are on a logarithmic scale and values for the rate of forcing result in time series of length 100, 200, 300, 400, and 500. Sample simulations for each model at different noise amplitude and rate of forcing are shown in Sup. Fig.~3. The transition time for each forced simulation is taken as the moment when the bifurcation parameter crosses the bifurcation, or the moment when the state variable crosses a threshold, if specified.\\

\noindent\textbf{(1) Fox model.} To test detection of a period-doubling bifurcation, we use a model of cardiac alternans \cite{fox2002period} with additive Gaussian white noise. This is given by
\begin{align}
    D_{n+1} & = (1-\alpha M_{n+1})\left(A + \frac{B}{1+e^{-(I_n-C)/D}}\right) + \sigma\epsilon_n,\\
    M_{n+1}& = e^{-I_n/\tau}[1+(M_n-1)e^{-D_n/\tau}],\\
    I_n &= T-D_n,
\end{align}
where $D_n$ is the action potential duration of the $n$th beat, $M_n$ is a memory variable, $I_n$ is the rest duration following the action potential, $T$ is the stimulation period, $\tau$ is the time constant of accumulation and dissipation of memory, $\alpha$ is the influence of memory on the action potential duration, and $A$, $B$, $C$ and $D$ are parameters governing the shape of the restitution curve. Following \cite{fox2002period}, we take $A=88$, $B=122$, $C=40$, $D=28$, $\tau=180$, $\alpha=0.2$, which give dynamics in good agreement with a complex ionic model. This yields a period-doubling bifurcation at approximately $T=200$. Forced simulations are run with $T$ decreasing linearly on the interval $[300, 150]$ and null simulations are run with $T=300$. 
Values for noise amplitude are $0.1 \times \{2^0, 2^{-1}, 2^{-2}, 2^{-3}, 2^{-4}\}$.\\

\noindent\textbf{(2) Westerhoff model.} To test detection of a Neimark-Sacker bifurcation, we use a simple model of business cycles based on consumer sentiment \cite{westerhoff2008consumer} with additive Gaussian noise. This is given by
    \begin{equation}
    Y_t = a +(b-d)Y_{t-1} + dY_{t-2} + \frac{cY_{t-1}}{1+\text{Exp}\left[-(Y_{t-1}-Y_{t-2})\right]} + \sigma \epsilon_t,
    \end{equation}
where $Y_t$ is the national income at time step $t$, $a$ is the level of autonomous expenditures of agents, $b$ and $c$ govern a curve that determines the fraction of income consumed by the agents, and $d$ is the policy-maker's control parameter to offset income trends. Following \cite{westerhoff2008consumer}, we take $b=0.45$, $c=0.1$, and $d=0$, which yields a Neimark-Sacker bifurcation at $a=20$ corresponding to the onset of business cycles. Forced simulations are run with $a$ increasing linearly on the interval $[10,22.5]$ and null simulations are run with $a=10$.
Values for noise amplitude are $0.1 \times \{2^0, 2^{-1}, 2^{-2}, 2^{-3}, 2^{-4}\}$.\\

\noindent\textbf{(3) Ricker model.} To test detection of a fold bifurcation, we use the Ricker model \cite{ricker1954stock} with a sigmoidal harvesting term and additive Gaussian noise. This is given by
    \begin{equation}
    x_{t+1} = x_t e^{r(1-x_t/k)} - F\frac{x_t^2}{x_t^2 + h^2} + \sigma \epsilon_t,
    \end{equation}
where $x_t$ is the population size at time step $t$, $r$ is the intrinsic growth rate, $k$ is the carrying capacity, $F$ is the harvesting rate, and $h$ governs the steepness of the sigmoidal harvesting term. We take $r=0.75$, $k=10$, $h=0.75$, which yields a fold bifurcation at $F=2.36$. Forced simulations are run with $F$ increasing linearly on the interval $[0,3.54]$ and null simulations are run with $F=0$. We define a transitions as the time when $x_t$ drops below 0.45. Values for noise amplitude are $0.2 \times \{2^0, 2^{-1}, 2^{-2}, 2^{-3}, 2^{-4}\}$.\\

\noindent\textbf{(4) Lotka-Volterra model.} To test detection of a transcritical bifurcation, we use the discrete-time analogue of the Lotka-Volterra model, first studied by Maynard Smith \cite{smith1968mathematical}. This system is especially relevant to arthropod predator-prey and host-parasitoid interactions. The rescaled equations \cite{neubert1992subcritical} are
\begin{align}
    x_{t+1} &= (r+1)x_t - rx_t^2 - cx_ty_t + \sigma \epsilon_t^{(1)} ,\\
    y_{t+1} & = cx_t y_t + \sigma \epsilon_t^{(2)},
\end{align}
$r$ relates to the growth rate of the prey ($x_t$), and $c$ relates to the foraging efficiency of the predator ($y_t$). We take $r=0.5$, which yields a transcritical bifurcation at $c=1$, the critical foraging efficiency beyond which the predator population can sustain themselves. Forced simulations are run with $c$ increasing linearly on the interval $[0.5,1.25]$ and null simulations are run with $c=0.5$. We look for early warning signals in the prey population. 
Values for noise amplitude are $0.01 \times \{2^0, 2^{-1}, 2^{-2}, 2^{-3}, 2^{-4}\}$.\\

\noindent\textbf{(5) Lorenz model.} To test detection of a pitchfork bifurcation, we use the reduced discrete Lorenz system, which was first introduced as a demonstration of computational chaos \cite{lorenz1989computational}. This is given by
\begin{align}
    x_{t+1} = (1+ah)x_t - h x_t y_t + \sigma \epsilon_t^{(1)} \\
    y_{t+1} = (1-h)y_t + h x_t ^2 + \sigma \epsilon_t^{(2)}
\end{align}
where state variables and parameters are derived from the full Lorenz equations \cite{lorenz1989computational}. We take $h=0.5$, which yields a pitchfork bifurcation at $a=0$. Forced simulations are run with $a$ increasing linearly over the interval $[-1, 0.25]$ and null simulations are run with $a=-1$. We look for early warning signals in $x_t$.
Values for noise amplitude are $0.01 \times \{2^0, 2^{-1}, 2^{-2}, 2^{-3}, 2^{-4}\}$.\\

\vspace{0.5cm}
\noindent\textbf{Experimental data}\\
We use data from spontaneously beating aggregates of chick heart cells, which can undergo a transition to an alternating rhythm with the addition of the potassium-channel blocker E-4031 \cite{kim2009stochastic}. To study the dynamics of these aggregates, we focus on the interbeat intervals---the time between consecutive beats. The transition to an alternating rhythm may be associated with a period-doubling bifurcation. Trends in variance and autocorrelation of the data have been shown to agree with theoretical expectations for a period-doubling bifurcation \cite{quail2015predicting}. We define the onset of the period-doubling bifurcation as the first time when the slope of a linear regression of the return map composed of a sliding window of interbeat intervals is below -0.95 for the next 10 beats. We have 23 time series that undergo a period-doubling bifurcation (Sup. Fig.~5, 6), and 18 time series that do not, from which we extract 23 segments at random with a random length between 100 and 500, to serve as null time series (Sup. Fig.~7, 8). For details on the chick heart experiments, see \cite{kim2009stochastic,quail2015predicting}.

\vspace{0.5cm}
\noindent\textbf{Computing and assessing the performance of EWS}\\
EWS are computed using the Python package \emph{ewstools} \cite{bury2023ewstools}. This involves first detrending the (pretransition) time series. For the model simulations, we use a Lowess filter with a span of 0.25 the length of the data. For the heart cell data, we use a Gaussian filter with a bandwidth of 20 beats. Variance and lag-1 autocorrelation are then computed over a rolling window of 0.5, which had higher performance than a rolling window of 0.25. The deep learning predictions at a given point in the time series are obtained by taking the preceding data from the time series, normalising it, prepending it with zeroes to make it 500 points in length, and feeding it into the classifier.

To compare performance of variance, lag-1 autocorrelation and the deep learning classifier, we use the AUC (area under curve) score of the ROC curve. The ROC curve plots the true positive rate vs. the false positive rate as a discrimination threshold is varied. We use the Kendall $\tau$ value as the discrimination threshold for variance and lag-1 autocorrelation, and the sum of the bifurcation probabilities for the discrimination threshold of the deep learning classifier.

\section*{Data availability}

Data from the chick heart aggregates, as well as synthetic data used to train and test the deep learning classifier are available at the GitHub repository \url{https://github.com/ThomasMBury/dl_discrete_bifurcation}.

\section*{Code availability}
Code and instructions to reproduce the analysis of this article are available at the GitHub repository \url{https://github.com/ThomasMBury/dl_discrete_bifurcation}. A reproducible run can be performed on Code Ocean at \url{https://codeocean.com/capsule/2209652/tree/v1} where the code is accompanied by the necessary software environment.

\section*{Acknowledgements}
This research was supported by a Fonds de Recherche du Québec – Nature et technologies (FRQNT) postdoctoral fellowship to T.M.B, a Canadian Institutes of Health Research (CIHR) grant (\#PJT-169008) to A.S., and compute services from the Digital Research Alliance of Canada (www.alliancecan.ca).

\printbibliography

@article{hastings2010regime,
	author = {Hastings, Alan and Wysham, Derin B},
	journal = {Ecology letters},
	number = {4},
	pages = {464--472},
	title = {Regime shifts in ecological systems can occur with no warning},
	volume = {13},
	year = {2010}}

@article{ditlevsen2010tipping,
	author = {Ditlevsen, Peter D and Johnsen, Sigfus J},
	journal = {Geophysical Research Letters},
	number = {19},
	title = {Tipping points: Early warning and wishful thinking},
	volume = {37},
	year = {2010}}

@article{boers2021observation,
	author = {Boers, Niklas},
	journal = {Nature Climate Change},
	number = {8},
	pages = {680--688},
	title = {Observation-based early-warning signals for a collapse of the Atlantic Meridional Overturning Circulation},
	volume = {11},
	year = {2021}}

@article{o2018stochasticity,
	author = {O'Regan, Suzanne M and Burton, Danielle L},
	journal = {Bulletin of Mathematical Biology},
	number = {6},
	pages = {1630--1654},
	title = {How stochasticity influences leading indicators of critical transitions},
	volume = {80},
	year = {2018}}

@article{ye2015equation,
	author = {Ye, Hao and Beamish, Richard J and Glaser, Sarah M and Grant, Sue CH and Hsieh, Chih-hao and Richards, Laura J and Schnute, Jon T and Sugihara, George},
	journal = {Proceedings of the National Academy of Sciences},
	number = {13},
	pages = {E1569--E1576},
	title = {Equation-free mechanistic ecosystem forecasting using empirical dynamic modeling},
	volume = {112},
	year = {2015}}

@article{barlow2014modelling,
	author = {Barlow, Lee-Ann and Cecile, Jacob and Bauch, Chris T and Anand, Madhur},
	journal = {PloS one},
	number = {4},
	pages = {e90511},
	title = {Modelling interactions between forest pest invasions and human decisions regarding firewood transport restrictions},
	volume = {9},
	year = {2014}}

@article{henderson2016alternative,
	author = {Henderson, Kirsten A and Bauch, Chris T and Anand, Madhur},
	journal = {Proceedings of the National Academy of Sciences},
	number = {51},
	pages = {14552--14559},
	title = {Alternative stable states and the sustainability of forests, grasslands, and agriculture},
	volume = {113},
	year = {2016}}

@article{quail2012chaotic,
	author = {Quail, Thomas and McVicar, Nevin and Aguilar, Martin and Kim, Min-Young and Hodge, Alex and Glass, Leon and Shrier, Alvin},
	journal = {Chaos: An Interdisciplinary Journal of Nonlinear Science},
	number = {3},
	pages = {033140},
	title = {Chaotic dynamics in cardiac aggregates induced by potassium channel block},
	volume = {22},
	year = {2012}}

@article{patel2023using,
	author = {Patel, Dhruvit and Ott, Edward},
	journal = {Chaos: An Interdisciplinary Journal of Nonlinear Science},
	number = {2},
	pages = {023143},
	title = {Using machine learning to anticipate tipping points and extrapolate to post-tipping dynamics of non-stationary dynamical systems},
	volume = {33},
	year = {2023}}

@article{kong2021machine,
	author = {Kong, Ling-Wei and Fan, Hua-Wei and Grebogi, Celso and Lai, Ying-Cheng},
	journal = {Physical Review Research},
	number = {1},
	pages = {013090},
	title = {Machine learning prediction of critical transition and system collapse},
	volume = {3},
	year = {2021}}

@article{bury2023ewstools,
	author = {Bury, Thomas M},
	journal = {Journal of Open Source Software},
	number = {82},
	pages = {5038},
	title = {ewstools: A {P}ython package for early warning signals of bifurcations in time series data},
	volume = {8},
	year = {2023}}

@article{wen2022transformers,
	author = {Wen, Qingsong and Zhou, Tian and Zhang, Chaoli and Chen, Weiqi and Ma, Ziqing and Yan, Junchi and Sun, Liang},
	journal = {arXiv},
	title = {Transformers in time series: A survey},
	year = {2022}}

@article{vaswani2017attention,
	author = {Vaswani, Ashish and Shazeer, Noam and Parmar, Niki and Uszkoreit, Jakob and Jones, Llion and Gomez, Aidan N and Kaiser, {\L}ukasz and Polosukhin, Illia},
	journal = {Advances in Neural Information Processing Systems},
	title = {Attention is all you need},
	volume = {30},
	year = {2017}}

@article{clements2019early,
	author = {Clements, Christopher F and McCarthy, Michael A and Blanchard, Julia L},
	journal = {Nature Communications},
	number = {1},
	pages = {1681},
	title = {Early warning signals of recovery in complex systems},
	volume = {10},
	year = {2019}}

@article{silla2011survey,
	author = {Silla, Carlos N and Freitas, Alex A},
	journal = {Data Mining and Knowledge Discovery},
	pages = {31--72},
	title = {A survey of hierarchical classification across different application domains},
	volume = {22},
	year = {2011}}

@article{lapeyrolerie2021teaching,
	author = {Lapeyrolerie, Marcus and Boettiger, Carl},
	journal = {Proceedings of the National Academy of Sciences},
	number = {40},
	pages = {e2115605118},
	title = {Teaching machines to anticipate catastrophes},
	volume = {118},
	year = {2021}}

@article{verrier2011microvolt,
	author = {Verrier, Richard L and Klingenheben, Thomas and Malik, Marek and El-Sherif, Nabil and Exner, Derek V and Hohnloser, Stefan H and Ikeda, Takanori and Mart{\'\i}nez, Juan Pablo and Narayan, Sanjiv M and Nieminen, Tuomo and others},
	journal = {Journal of the American College of Cardiology},
	number = {13},
	pages = {1309--1324},
	title = {Microvolt T-wave alternans: physiological basis, methods of measurement, and clinical utility---consensus guideline by International Society for Holter and Noninvasive Electrocardiology},
	volume = {58},
	year = {2011}}

@article{kuehn2013mathematical,
	author = {Kuehn, Christian},
	journal = {Journal of Nonlinear Science},
	number = {3},
	pages = {457--510},
	title = {A mathematical framework for critical transitions: normal forms, variance and applications},
	volume = {23},
	year = {2013}}

@book{scheffer2020critical,
	author = {Scheffer, Marten},
	publisher = {Princeton University Press},
	title = {Critical transitions in nature and society},
	year = {2020}}

@book{sornette2017stock,
	author = {Sornette, Didier},
	publisher = {Princeton University Press},
	title = {Why Stock Markets Crash},
	year = {2017}}

@article{boers2018early,
	author = {Boers, Niklas},
	journal = {Nature Communications},
	number = {1},
	pages = {2556},
	title = {Early-warning signals for {D}ansgaard-{O}eschger events in a high-resolution ice core record},
	volume = {9},
	year = {2018}}

@article{wang2012flickering,
  title={Flickering gives early warning signals of a critical transition to a eutrophic lake state},
  author={Wang, Rong and Dearing, John A and Langdon, Peter G and Zhang, Enlou and Yang, Xiangdong and Dakos, Vasilis and Scheffer, Marten},
  journal={Nature},
  volume={492},
  number={7429},
  pages={419--422},
  year={2012},
  publisher={Nature Publishing Group UK London}
}

@article{hagstrom2021phase,
  author = {Hagstrom, George I. and Levin, Simon A.},
  journal = {arXiv},
  title = {Phase Transitions and the Theory of Early Warning Indicators for Critical Transitions},  
  year = {2021},
}

@article{levin1998ecosystems,
  title={Ecosystems and the biosphere as complex adaptive systems},
  author={Levin, Simon A},
  journal={Ecosystems},
  volume={1},
  pages={431--436},
  year={1998},
  publisher={Springer}
}

@article{hennekam2020early,
	author = {Hennekam, Rick and van der Bolt, Bregje and van Nes, Egbert H and de Lange, Gert J and Scheffer, Marten and Reichart, Gert-Jan},
	journal = {Geophysical Research Letters},
	number = {20},
	pages = {e2020GL089183},
	title = {Early-warning signals for marine anoxic events},
	volume = {47},
	year = {2020}}

@article{pavithran2021effect,
	author = {Pavithran, Induja and Sujith, RI},
	journal = {Chaos: An Interdisciplinary Journal of Nonlinear Science},
	number = {1},
	pages = {013116},
	title = {Effect of rate of change of parameter on early warning signals for critical transitions},
	volume = {31},
	year = {2021}}

@article{pace2017reversal,
	author = {Pace, Michael L and Batt, Ryan D and Buelo, Cal D and Carpenter, Stephen R and Cole, Jonathan J and Kurtzweil, Jason T and Wilkinson, Grace M},
	journal = {Proceedings of the National Academy of Sciences},
	pages = {201612424},
	title = {Reversal of a cyanobacterial bloom in response to early warnings},
	year = {2016}}

@article{dakos2008slowing,
	author = {Dakos, Vasilis and Scheffer, Marten and van Nes, Egbert H and Brovkin, Victor and Petoukhov, Vladimir and Held, Hermann},
	journal = {Proceedings of the National Academy of Sciences},
	number = {38},
	pages = {14308--14312},
	title = {Slowing down as an early warning signal for abrupt climate change},
	volume = {105},
	year = {2008}}

@article{kefi2013early,
	author = {K{\'e}fi, Sonia and Dakos, Vasilis and Scheffer, Marten and Van Nes, Egbert H and Rietkerk, Max},
	journal = {Oikos},
	number = {5},
	pages = {641--648},
	title = {Early warning signals also precede non-catastrophic transitions},
	volume = {122},
	year = {2013}}

@article{wissel1984universal,
	author = {Wissel, C},
	journal = {Oecologia},
	number = {1},
	pages = {101--107},
	title = {A universal law of the characteristic return time near thresholds},
	volume = {65},
	year = {1984}}

@article{clements2016rate,
	author = {Clements, Christopher F and Ozgul, Arpat},
	journal = {Ecology and Evolution},
	number = {21},
	pages = {7787--7793},
	title = {Rate of forcing and the forecastability of critical transitions},
	volume = {6},
	year = {2016}}

@article{wiesenfeld1985noisy,
	author = {Wiesenfeld, Kurt},
	journal = {Journal of Statistical Physics},
	number = {5},
	pages = {1071--1097},
	title = {Noisy precursors of nonlinear instabilities},
	volume = {38},
	year = {1985}}

@article{deb2022machine,
	author = {Deb, Smita and Sidheekh, Sahil and Clements, Christopher F and Krishnan, Narayanan C and Dutta, Partha S},
	journal = {Royal Society Open Science},
	number = {2},
	pages = {211475},
	title = {Machine learning methods trained on simple models can predict critical transitions in complex natural systems},
	volume = {9},
	year = {2022}}

@article{scheffer2009early,
	author = {Scheffer, Marten and Bascompte, Jordi and Brock, William A and Brovkin, Victor and Carpenter, Stephen R and Dakos, Vasilis and Held, Hermann and Van Nes, Egbert H and Rietkerk, Max and Sugihara, George},
	journal = {Nature},
	number = {7260},
	pages = {53--59},
	title = {Early-warning signals for critical transitions},
	volume = {461},
	year = {2009}}

@article{may1974biological,
	author = {May, Robert M},
	journal = {Science},
	number = {4164},
	pages = {645--647},
	title = {Biological populations with nonoverlapping generations: stable points, stable cycles, and chaos},
	volume = {186},
	year = {1974}}

@book{strogatz2018nonlinear,
	author = {Strogatz, Steven H},
	publisher = {CRC press},
	title = {Nonlinear Dynamics and Chaos: with Applications to Physics, Biology, Chemistry, and Engineering},
	year = {2018}}

@book{glass2020clocks,
	author = {Glass, Leon and Mackey, Michael C},
	publisher = {Princeton University Press},
	title = {From Clocks to Chaos},
	year = {2020}}

@article{allen1994some,
	author = {Allen, Linda JS},
	journal = {Mathematical Biosciences},
	number = {1},
	pages = {83--105},
	title = {Some discrete-time {SI}, {SIR}, and {SIS} epidemic models},
	volume = {124},
	year = {1994}}

@article{bury2020detecting,
	author = {Bury, Thomas M and Bauch, Chris T and Anand, Madhur},
	journal = {Journal of the Royal Society Interface},
	number = {170},
	pages = {20200482},
	title = {Detecting and distinguishing tipping points using spectral early warning signals},
	volume = {17},
	year = {2020}}

@article{bury2021deep,
	author = {Bury, Thomas M and Sujith, RI and Pavithran, Induja and Scheffer, Marten and Lenton, Timothy M and Anand, Madhur and Bauch, Chris T},
	journal = {Proceedings of the National Academy of Sciences},
	number = {39},
	pages = {e2106140118},
	title = {Deep learning for early warning signals of tipping points},
	volume = {118},
	year = {2021}}

@book{kuznetsov1998elements,
	author = {Kuznetsov, Yuri},
	publisher = {Springer},
	title = {Elements of Applied Bifurcation Theory},
	year = {1998}}

@book{smith1968mathematical,
	author = {Smith, J Maynard},
	publisher = {CUP Archive},
	title = {Mathematical Ideas in Biology},
	year = {1968}}

@article{ricker1954stock,
	author = {Ricker, William Edwin},
	journal = {Journal of the Fisheries Board of Canada},
	number = {5},
	pages = {559--623},
	title = {Stock and recruitment},
	volume = {11},
	year = {1954}}

@article{fox2002period,
	author = {Fox, Jeffrey J and Bodenschatz, Eberhard and Gilmour Jr, Robert F},
	journal = {Physical Review Letters},
	number = {13},
	pages = {138101},
	title = {Period-doubling instability and memory in cardiac tissue},
	volume = {89},
	year = {2002}}

@article{quail2015predicting,
	author = {Quail, Thomas and Shrier, Alvin and Glass, Leon},
	journal = {Proceedings of the National Academy of Sciences},
	number = {30},
	pages = {9358--9363},
	title = {Predicting the onset of period-doubling bifurcations in noisy cardiac systems},
	volume = {112},
	year = {2015}}

@article{kim2009stochastic,
	author = {Kim, Min-Young and Aguilar, Martin and Hodge, Alex and Vigmond, Edward and Shrier, Alvin and Glass, Leon},
	journal = {Physical Review Letters},
	number = {5},
	pages = {058101},
	title = {Stochastic and spatial influences on drug-induced bifurcations in cardiac tissue culture},
	volume = {103},
	year = {2009}}

@article{dylewsky2022universal,
	author = {Dylewsky, Daniel and Lenton, Timothy M and Scheffer, Marten and Bury, Thomas M and Fletcher, Christopher G and Anand, Madhur and Bauch, Chris T},
	journal = {arXiv},
	title = {Universal Early Warning Signals of Phase Transitions in Climate Systems},
	year = {2022}}

@article{lorenz1989computational,
	author = {Lorenz, Edward N},
	journal = {Physica D: Nonlinear Phenomena},
	number = {3},
	pages = {299--317},
	title = {Computational chaos-a prelude to computational instability},
	volume = {35},
	year = {1989}}

@article{neubert1992subcritical,
	author = {Neubert, Michael G and Kot, Mark},
	journal = {Mathematical Biosciences},
	number = {1},
	pages = {45--66},
	title = {The subcritical collapse of predator populations in discrete-time predator-prey models},
	volume = {110},
	year = {1992}}

@article{westerhoff2008consumer,
	author = {Westerhoff, Frank H},
	journal = {Applied Economics Letters},
	number = {15},
	pages = {1201--1205},
	title = {Consumer sentiment and business cycles: a {N}eimark--{S}acker bifurcation scenario},
	volume = {15},
	year = {2008}}

\end{document}


\maketitle

\begin{figure}[ht]
    \centering
    \includegraphics[width=0.7\textwidth]{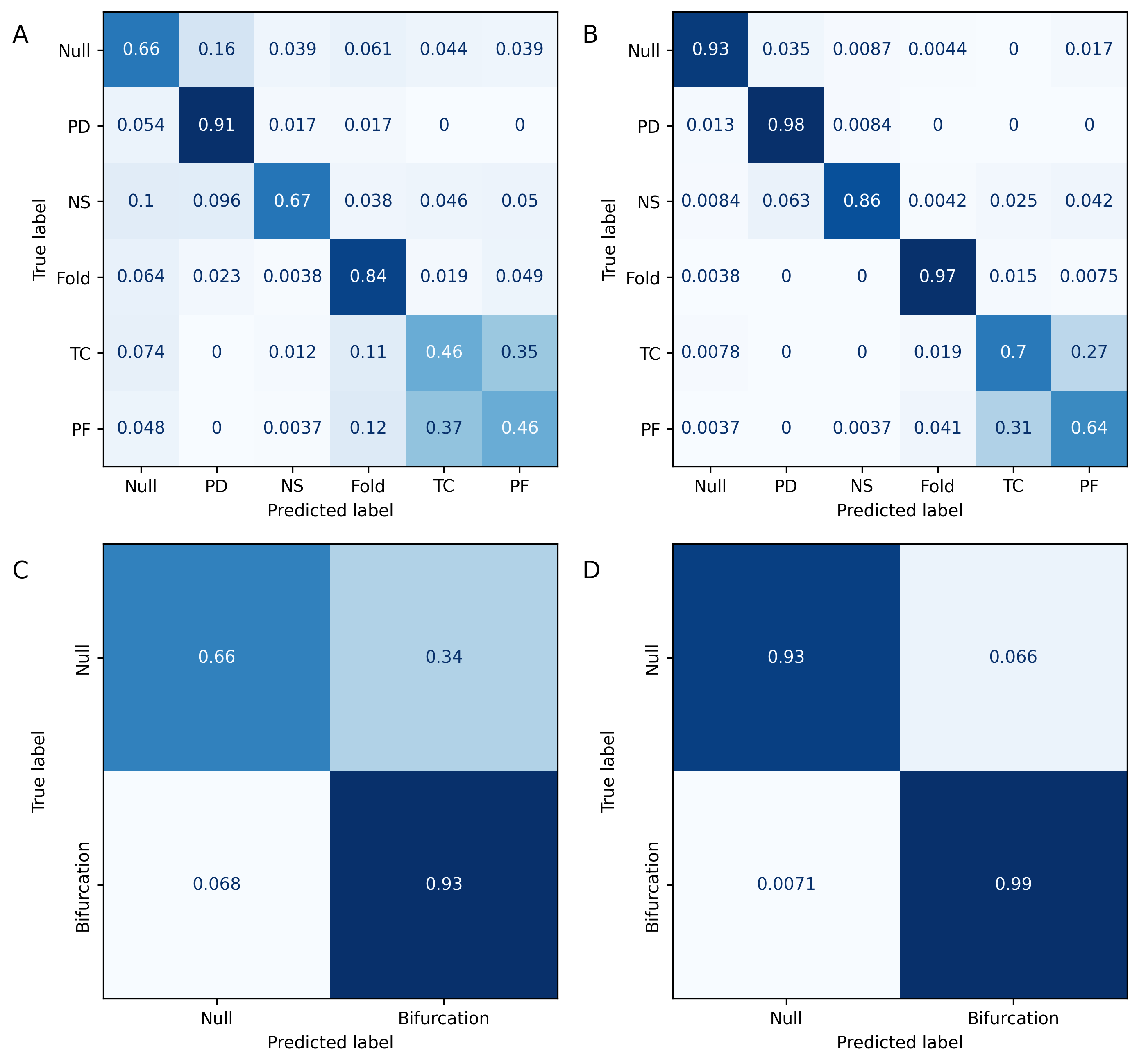}
    \caption{Confusion matrices summarising the performance of Classifier 1 and Classifier 2 on their respective test sets for the multi-class and binary classification problem. Cell values show (row-)normalised classification rates for each class. (A) Classifier 1 on the multi-class classification problem obtains an F1 score of $0.66$. (B) Classifier 2 on the multi-class classification problem obtains an F1 score of $0.85$. (C) Classifier 1 on the binary classification problem obtains an F1 score of $0.82$. (D) Classifier 2 on the binary classification problem obtains an F1 score of $0.98$. PD: period-doubling. NS: Neimark-Sacker. TC: transcritical. PF: pitchfork.}
    \label{fig:confusion_matrices}
\end{figure}

\begin{figure}[ht]
    \centering
    \includegraphics[width=0.5\textwidth]{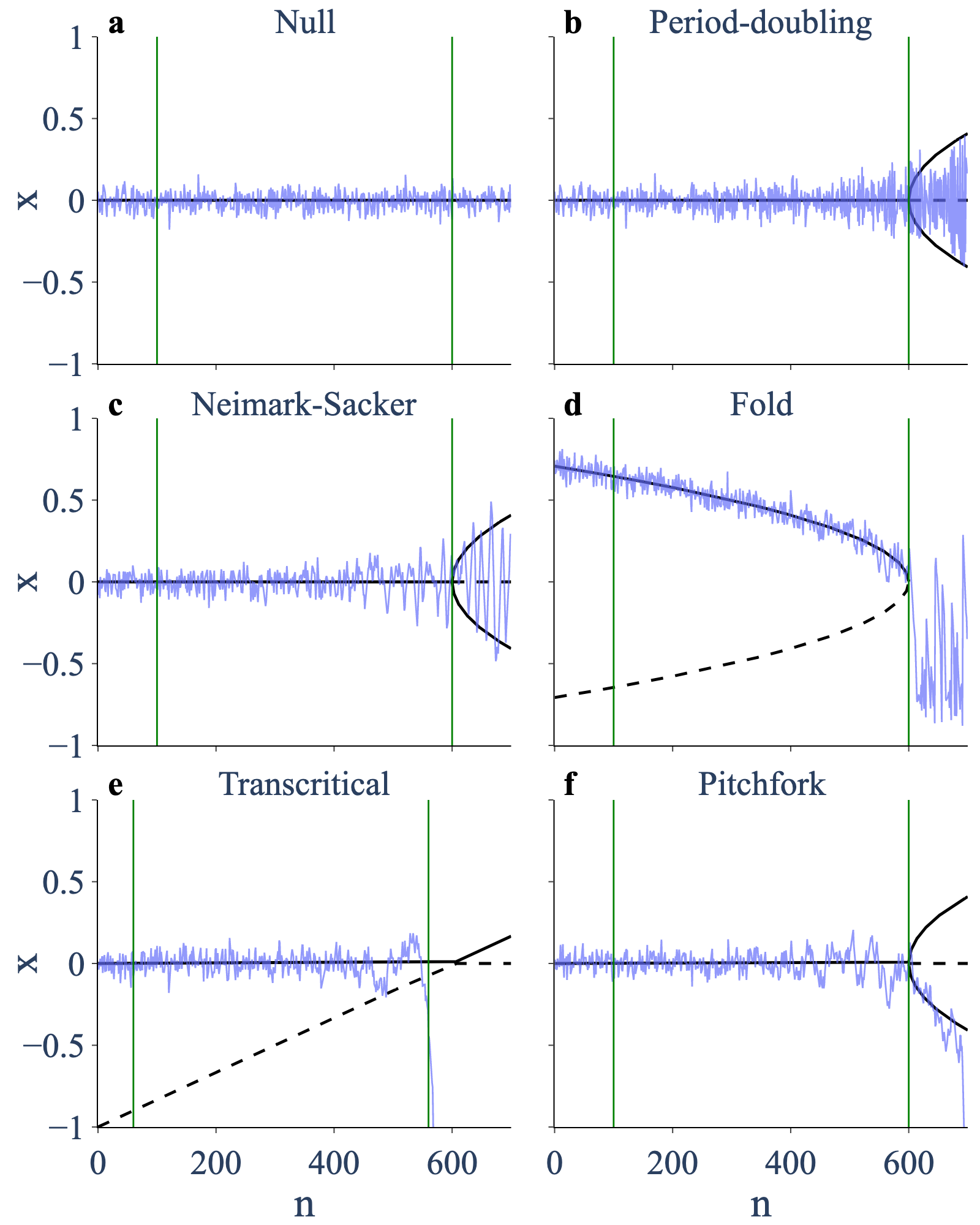}
    \caption{Sample simulations and corresponding bifurcation diagrams for each class in the training data. Panels show the trajectory of $x_n$ (blue), the corresponding bifurcation diagram (black) and the section used for training in each case (between the green vertical lines). The six possible classes are (a) null, (b) period-doubling, (c) Neimark-Sacker, (d) fold, (e) transcritical, and (f) pitchfork. These samples show supercritical versions of the period-doubling, Neimark-Sacker and pitchfork bifurcations, though subcritical examples are also present in the training data. The end time for the extracted data is taken as the time the bifurcation is crossed ($t=600$) or the first time the deviation from equilibrium reaches ten times the noise amplitude, whichever comes first.}
    \label{fig:training_samples}
\end{figure}

\begin{figure}[ht]
    \centering
    \includegraphics[width=0.9\textwidth]{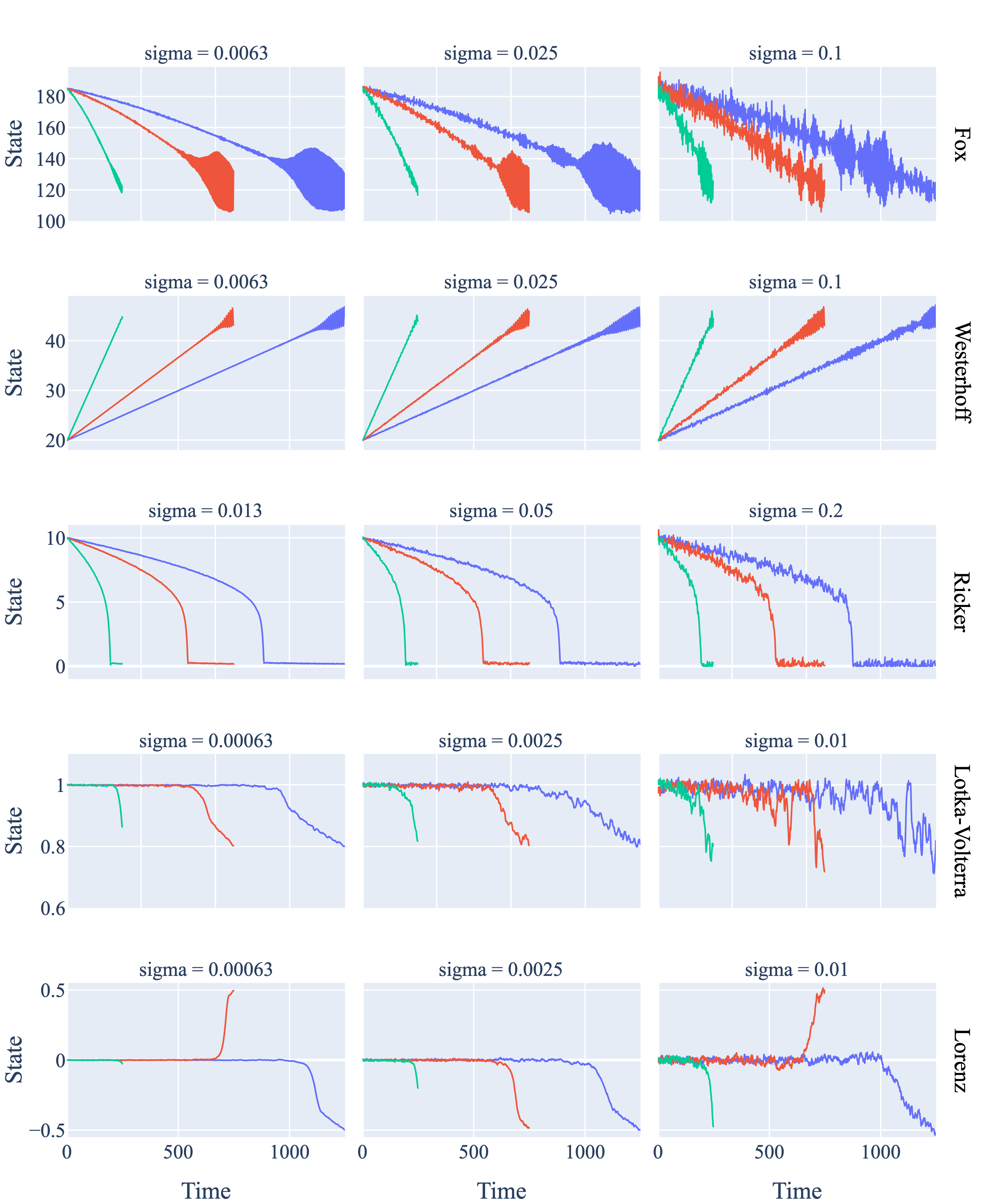}
    \caption{Sample simulations of each theoretical model approaching a bifurcation under a range of noise amplitudes (sigma) and rates of forcing (colour). The rates of forcing are chosen such that the bifurcation is crossed after 100 (green), 300 (red) and 500 (blue) units of time. The Fox model goes through a period-doubling bifurcation, the Westerhoff model a Neimark-Sacker bifurcation, the Ricker model a fold bifurcation, the Lotka-Volterra model a transcritical bifurcation, and the Lorenz model a pitchfork bifurcation. Model parameters are provided in Methods.}
    \label{fig:model_samples}
\end{figure}

\begin{figure}[ht]
    \centering
    \includegraphics[width=0.9\textwidth]{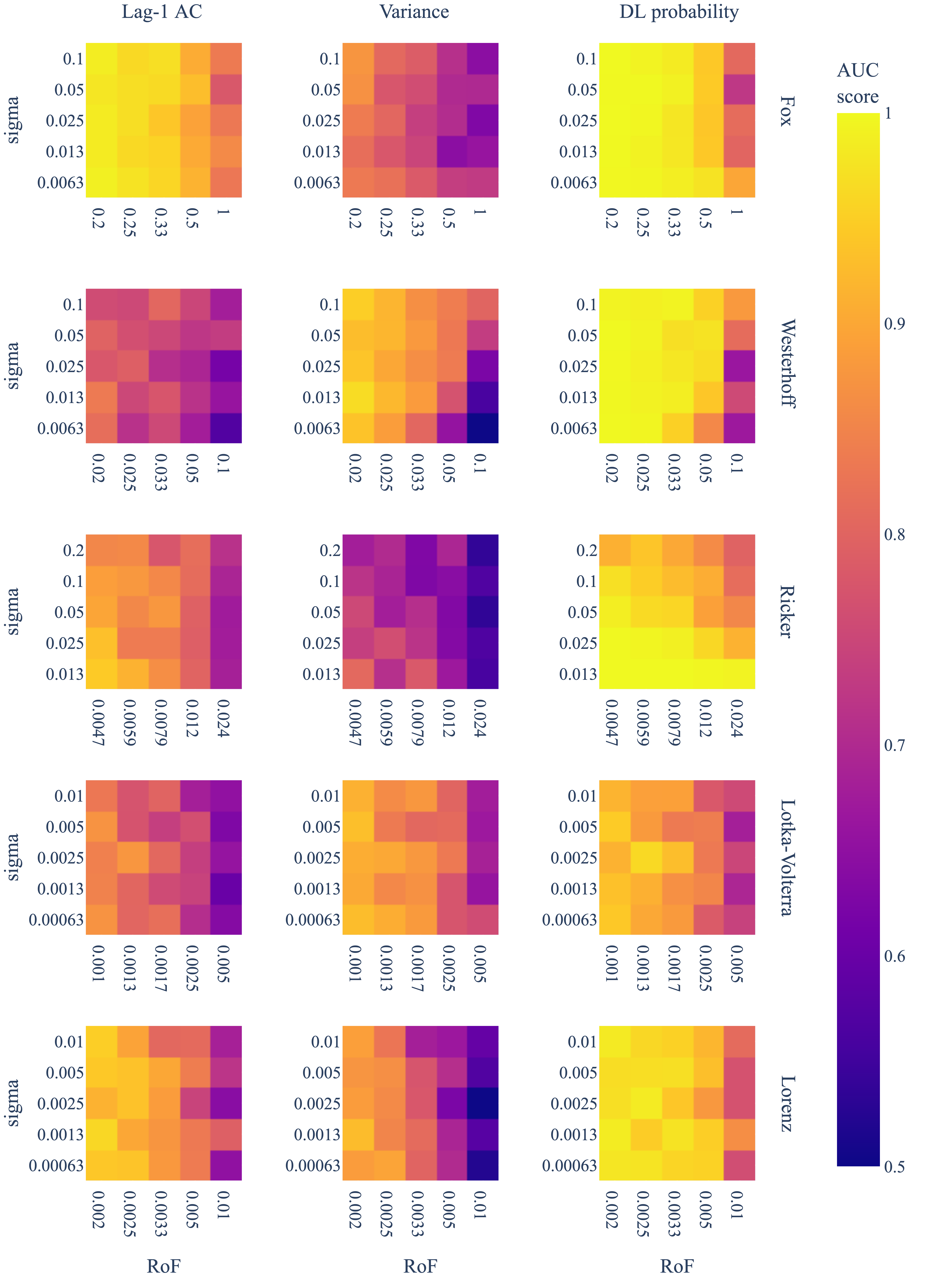}
    \caption{AUC score (area under the ROC curve) at different values of noise amplitude (sigma) and rate of forcing (RoF) for each indicator (columns) and theoretical model (rows). At each combination of RoF and sigma, we run 100 forced and null simulations, resulting in a total of 5000 simulations for each model. ROC curves are computed using predictions at $80\%$ of the way through the pretransition time series. The bifurcations associated with each model are period-doubling (Fox), Neimark-Sacker (Westerhoff), fold (Ricker), transcritical (Lotka-Volterra), and pitchfork (Lorenz).}
    \label{fig:auc_rof_sigma}
\end{figure}

\begin{figure}[ht]
    \centering
    \includegraphics[width=0.4\textwidth]{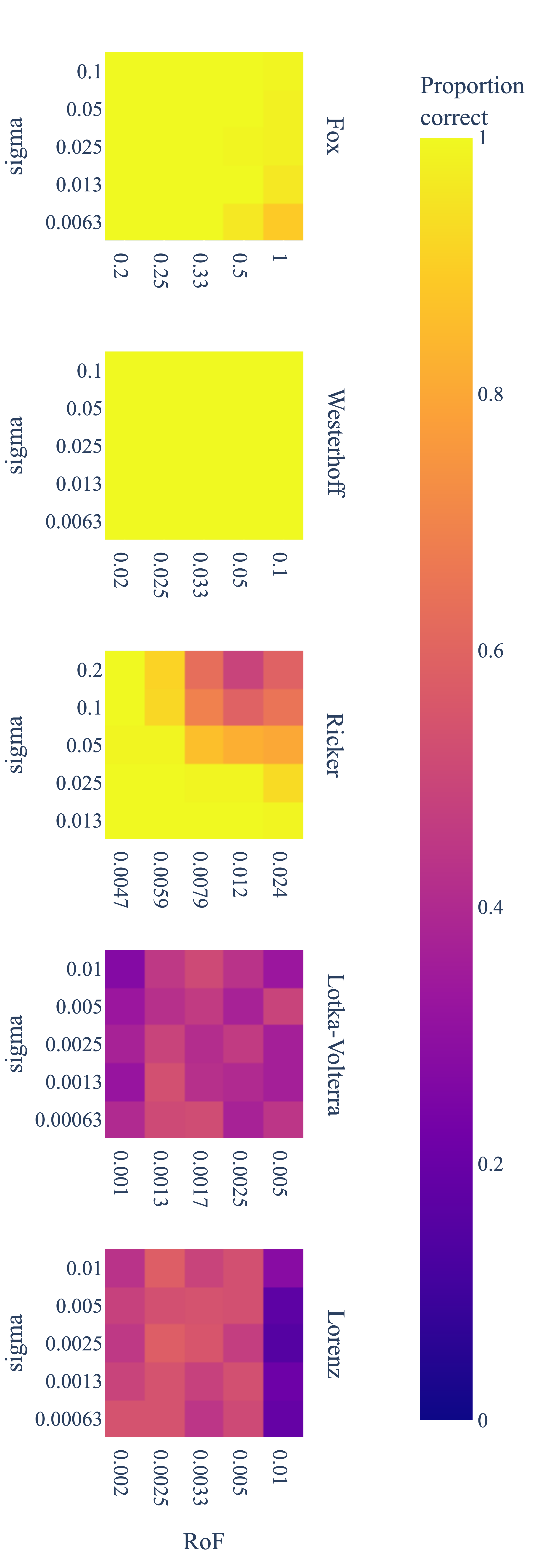}
    \caption{Proportion of predictions from the classifier that favoured the correct bifurcation for forced trajectories of different noise amplitude (sigma) and rate of forcing (RoF). For each combination of RoF and sigma, 100 forced trajectories are simulated, and predictions are made 80\% of the way through the pretransition time series. The bifurcations associated with each model are period-doubling (Fox), Neimark-Sacker (Westerhoff), fold (Ricker), transcritical (Lotka-Volterra), and pitchfork (Lorenz).}
    \label{fig:dl_rof_sigma}
\end{figure}

\begin{figure}[ht!]
    \centering
    \includegraphics[width=\textwidth]{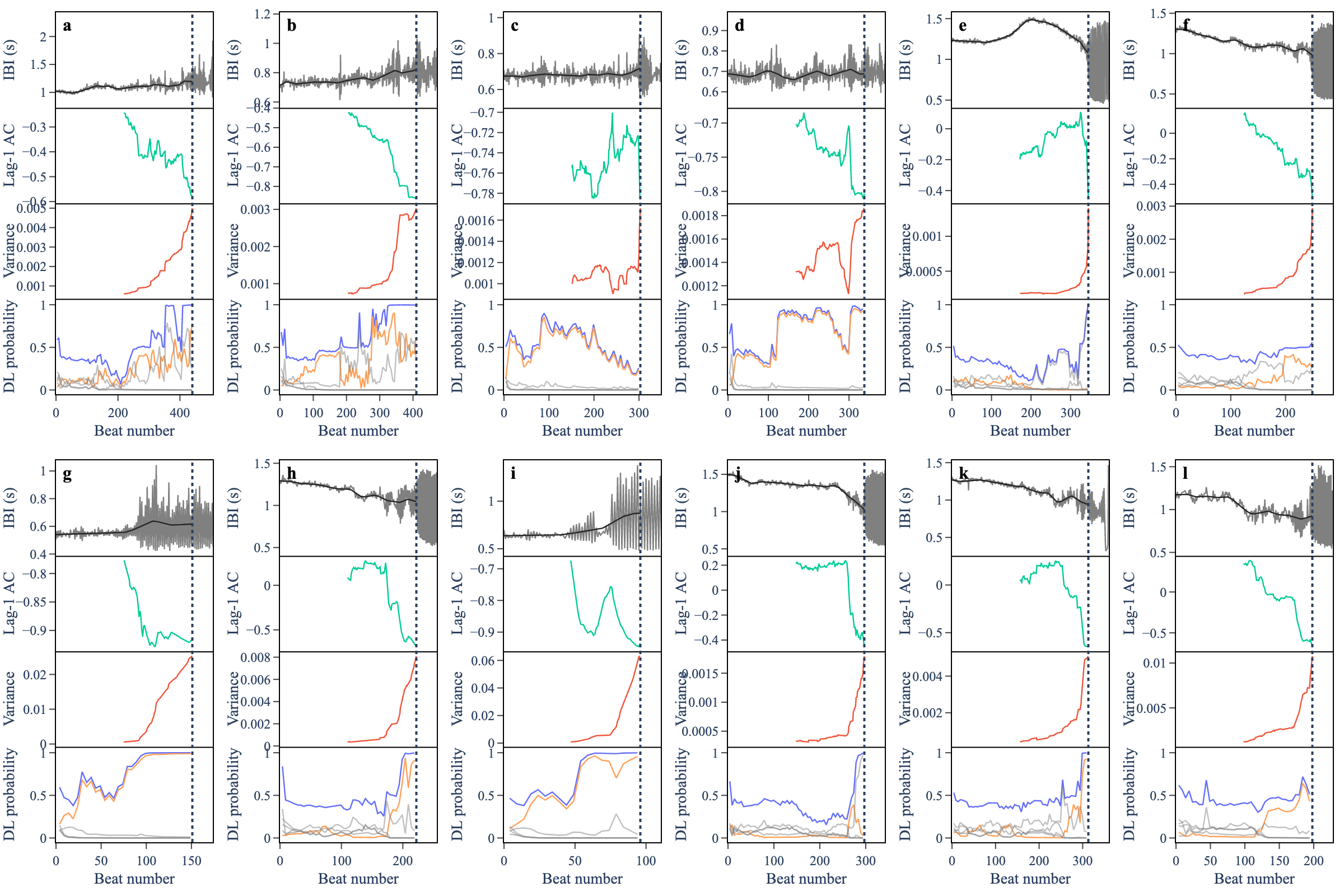}
    \caption{Trends in indicators prior to a period doubling bifurcation in chick heart aggregates (a-l). See Fig.~3 caption for details.}
    \label{fig:chick_pd_1}
\end{figure}

\begin{figure}[ht!]
    \centering
    \includegraphics[width=\textwidth]{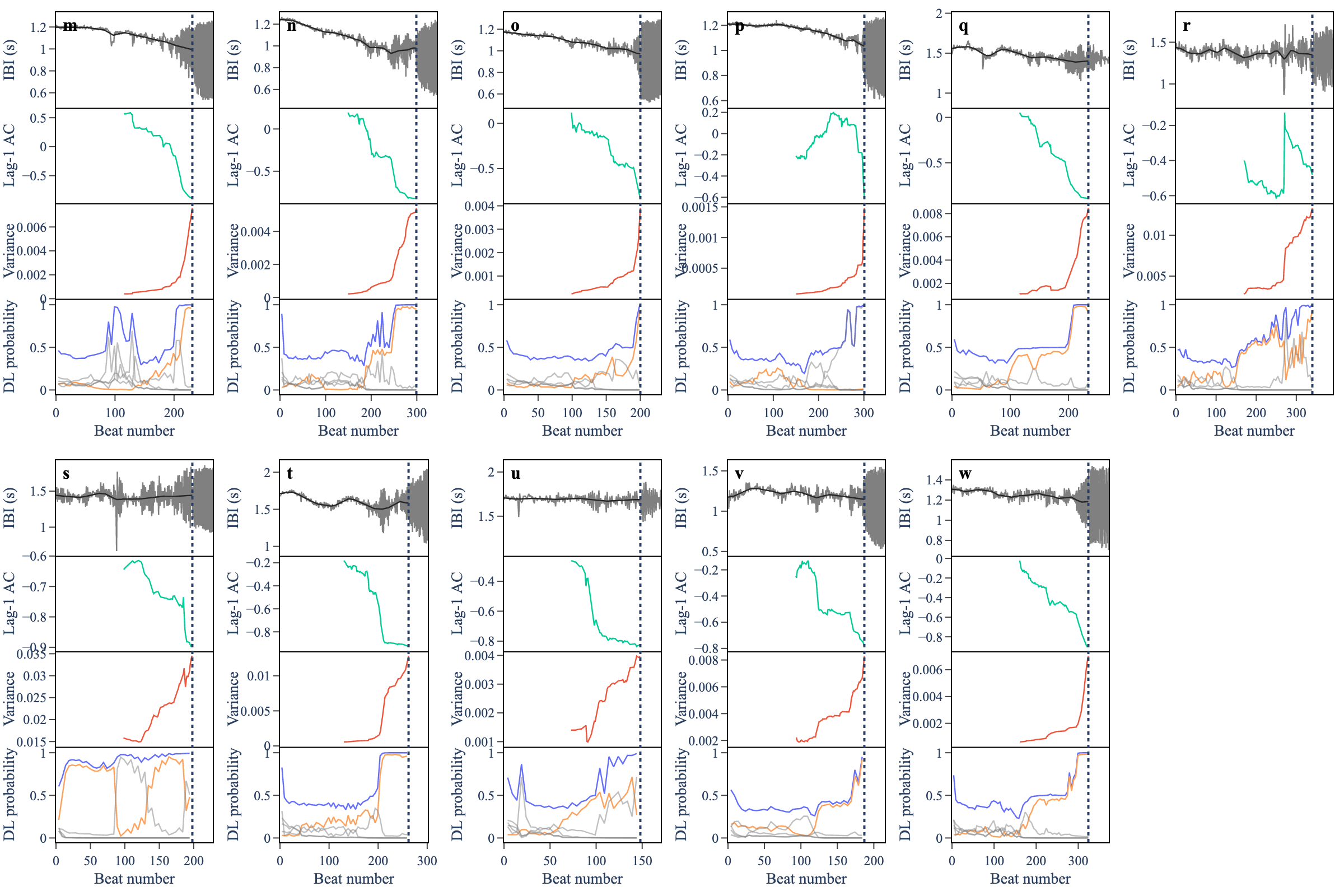}
    \caption{Trends in indicators prior to a period doubling bifurcation in chick heart aggregates (m-w). See Fig.~3 caption for details.}
    \label{fig:chick_pd_2}
\end{figure}

\begin{figure}[ht!]
    \centering
    \includegraphics[width=\textwidth]{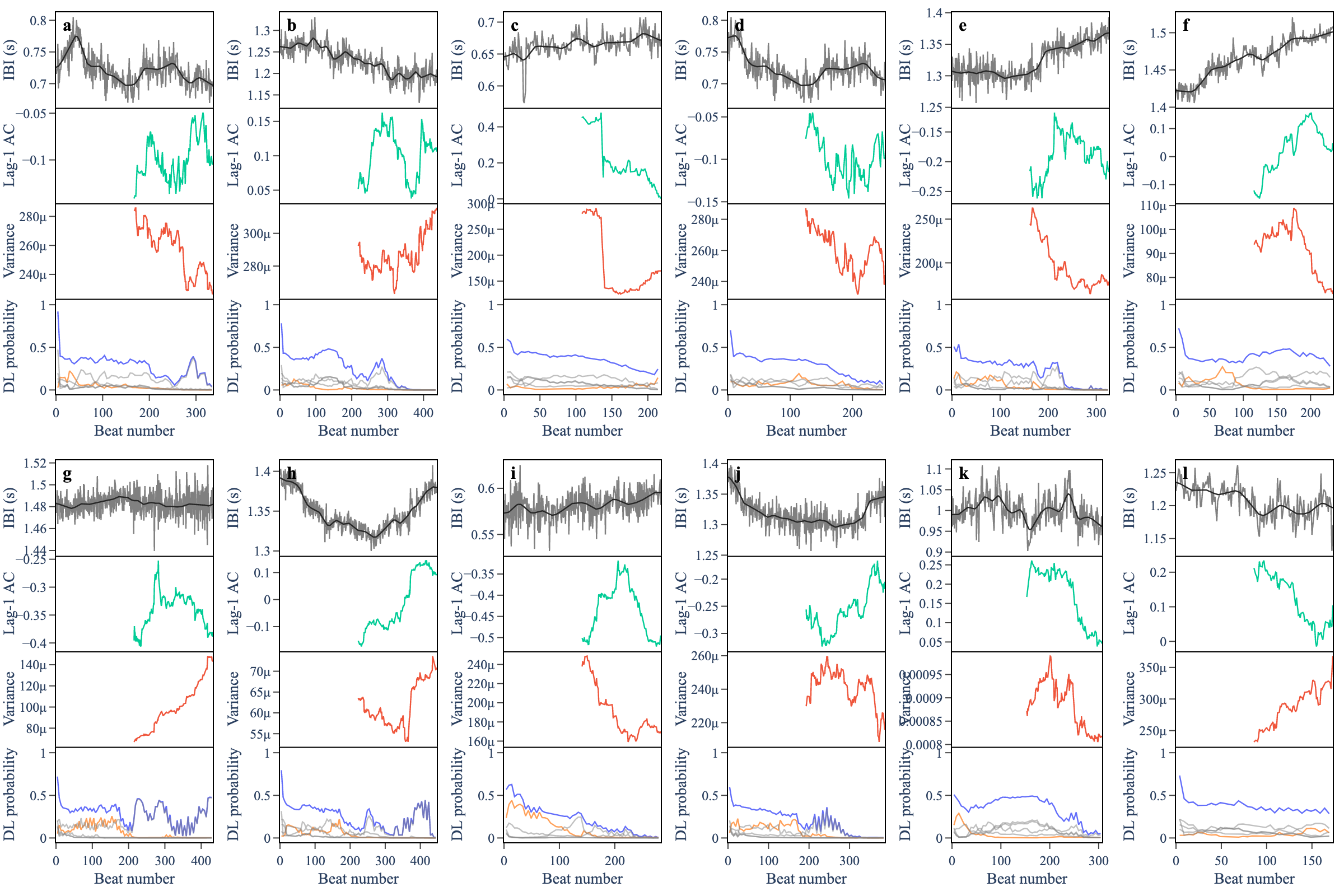}
    \caption{Trends in indicators for chick heart aggregates that do not undergo a period-doubling bifurcation (a-l). See Fig.~3 caption for details.}
    \label{fig:chick_null_1}
\end{figure}

\begin{figure}[ht!]
    \centering
    \includegraphics[width=\textwidth]{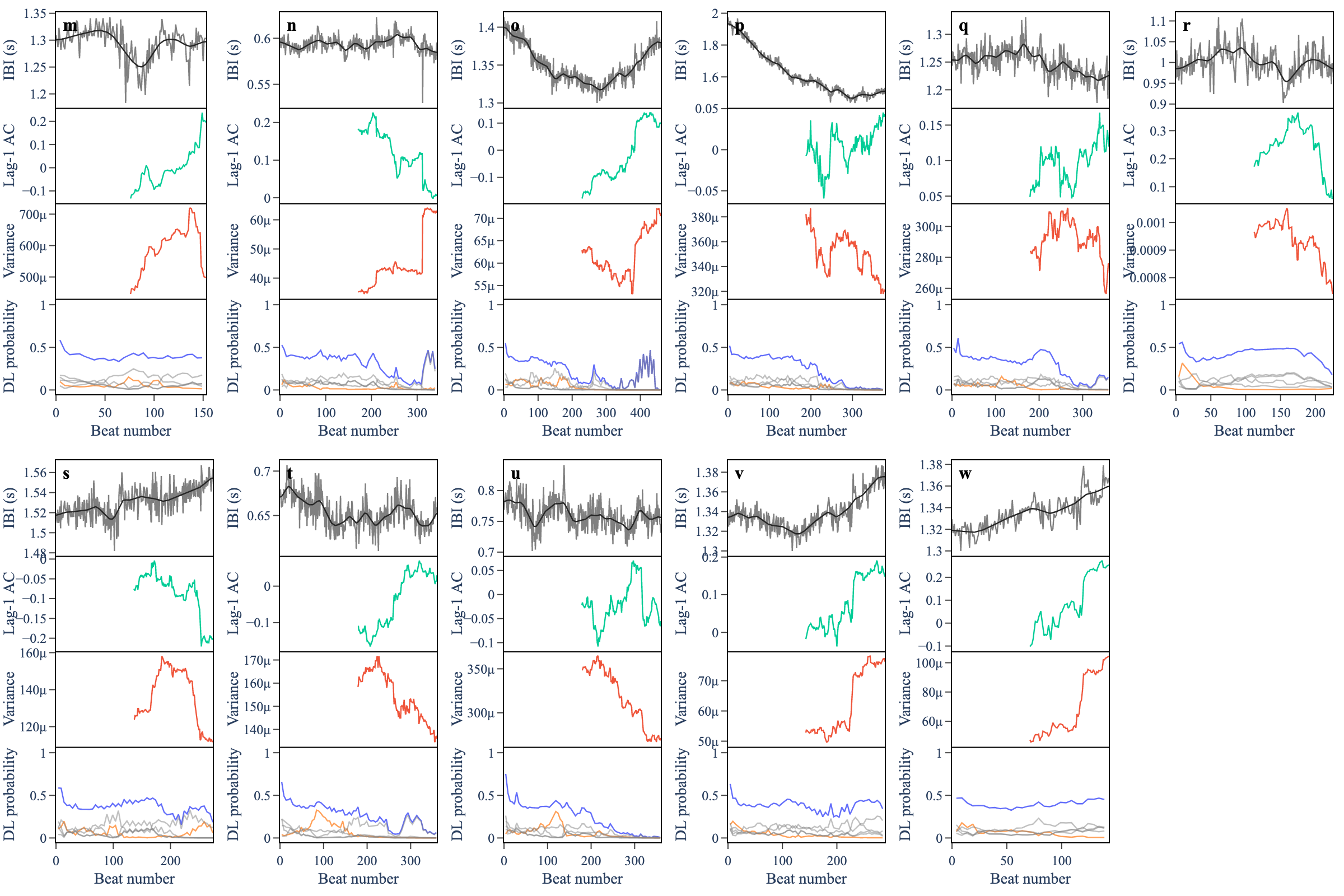}
    \caption{Trends in indicators for chick heart aggregates that do not undergo a period-doubling bifurcation (m-w). See Fig.~3 caption for details.}
    \label{fig:chick_null_2}
\end{figure}